\documentclass[prd,preprintnumbers,nofootinbib,superscriptaddress,onecolumn,preprint]{revtex4}

\pdfoutput=1
\usepackage{graphicx}
\usepackage{amsmath}
\usepackage{amssymb}
\usepackage{slashed}
\usepackage{mathtools}
\usepackage[utf8]{inputenc}
\usepackage[colorlinks=true,linkcolor=blue,bookmarksopen,bookmarksnumbered]{hyperref}
\usepackage{color}
\usepackage[usenames,dvipsnames]{xcolor}
\usepackage[normalem]{ulem}
\usepackage{soul}
\usepackage{units}
\usepackage{rotating}
\usepackage{hhline,multirow,tabularx}

\newcommand{\T}[1]{\boldsymbol{#1}_{\text{T}}}
\newcommand{\kT}{\ensuremath{k_{\rm T}}}
\newcommand{\ktmax}{\ensuremath{k_{\rm T max}}}
\newcommand{\ktmaxsq}{\ensuremath{k_{\rm T max}^2}}
\newcommand\3[1]{\boldsymbol{#1}}
\newcommand{\Tsc}[2]{#1_{#2\text{T}}}
\newcommand{\Tscsq}[2]{#1^2_{#2\text{T}}}
\newcommand{\Tscsqb}[2]{\boldsymbol{#1}^2_{#2\text{T}}}
\newcommand{\no}{\nonumber \\}
\newcommand{\parz}[1]{\ensuremath{\left(#1\right)}}
\newcommand{\order}[1]{\ensuremath{O\parz{#1}}}
\newcommand{\mhad}{\ensuremath{M}}
\newcommand{\alfa}{\ensuremath{\alpha}}
\newcommand{\xbj}{\ensuremath{x_{\rm bj}}}
\newcommand{\xn}{\ensuremath{x_{\rm n}}}
\newcommand{\mquark}{\ensuremath{m_{q}}}
\newcommand{\mgluon}{\ensuremath{m_{s}}}
\newcommand{\spectator}{\ensuremath{p_s}}
\newcommand{\jet}{\ensuremath{p_q}}

\newcommand{\prop}{{[ \ensuremath{\rm Prop} ]}}
\newcommand{\num}{{\ensuremath{\rm T}}}
\newcommand{\jac}[1]{{\ensuremath{[ {\rm Jac_{#1}} ] }}}

\newcommand{\diff}[1]{\mathrm{d}#1}

\newcommand{\eref}[1]{Eq.~(\ref{e.#1})}
\newcommand{\erefs}[2]{Eqs.~(\ref{e.#1})--(\ref{e.#2})}
\newcommand{\fref}[1]{Fig.~\ref{f.#1}}

\newcommand{\sref}[1]{Sec.~\ref{s.#1}}

\begin{document}

\title{What are the low-$Q$ and large-$x$ boundaries of \\
	collinear QCD factorization theorems?}

\preprint{JLAB-THY-17-2413}
\author{E.~Moffat}
\email{emoff003@odu.edu}
\affiliation{Department of Physics, Old Dominion University, Norfolk, VA 23529, USA}
\author{W.~Melnitchouk}
\email{wmelnitc@jlab.org}
\affiliation{Jefferson Lab, 12000 Jefferson Avenue, Newport News, VA 23606, USA}
\author{T.~C.~Rogers}
\email{tedconantrogers@gmail.com}
\affiliation{Department of Physics, Old Dominion University, Norfolk, VA 23529, USA}
\affiliation{Jefferson Lab, 12000 Jefferson Avenue, Newport News, VA 23606, USA}
\author{N.~Sato}
\email{nsato@jlab.org}
\affiliation{Jefferson Lab, 12000 Jefferson Avenue, Newport News, VA 23606, USA}

\date{\today}

%%%%%%%%%%%%%%%%%%%%%%%%%%%%%%%%%%%%%%%%%%%%%%%%%%%%%%%%%%%%%%%%%%%%%%%%
\begin{abstract}

Familiar factorized descriptions of classic QCD processes such as
deeply-inelastic scattering (DIS) apply in the limit of very large
hard scales, much larger than nonperturbative mass scales and other
nonperturbative physical properties like intrinsic transverse momentum.
Since many interesting DIS studies occur at kinematic regions where
the hard scale, $Q \sim$~1--2~GeV, is not very much greater than the
hadron masses involved, and the Bjorken scaling variable $\xbj$ is
large, $\xbj \gtrsim 0.5$, it is important to examine the boundaries of
the most basic factorization assumptions and assess whether improved
starting points are needed.
Using an idealized field-theoretic model that contains most of the
essential elements that a factorization derivation must confront,
we retrace the steps of factorization approximations and compare
with calculations that keep all kinematics exact.
We examine the relative importance of such quantities as the target
mass, light quark masses, and intrinsic parton transverse momentum,
and argue that a careful accounting of parton virtuality is essential
for treating power corrections to collinear factorization.
We use our observations to motivate searches for new or enhanced factorization
theorems specifically designed to deal with moderately low-$Q$ 
and large-$\xbj$ physics.

\end{abstract}

\maketitle

%%%%%%%%%%%%%%%%%%%%%%%%%%%%%%%%%%%%%%%%%%%%%%%%%%%%%%%%%%%%%%%%%%%%%%%%
\section{Introduction}
\label{s.intro}

Factorization theorems deal with the way interactions at different
spacetime scales disentangle, for certain classes of scattering
processes, in the asymptotically large limit of some physical energy
\cite{Collins:1988gx}.  They are especially important in QCD where
asymptotic freedom enables calculations of short-distance partonic
amplitudes using small-coupling perturbation theory.  Many interesting
applications of QCD factorization in hadronic physics are in regions
where small-coupling techniques are likely to be useful, but where
familiar kinematical approximations are perhaps questionable, and
where the interplay between perturbative and nonperturbative physics
becomes more intricate than at the very highest available energies.

Deeply-inelastic scattering (DIS) of leptons from hadrons at moderately
low momentum transfers $Q$ is a prototypical example of this.
Scales of $Q \sim$~1--2~GeV correspond to $\alpha_s/\pi \lesssim 0.1$,
where $\alpha_s$ is the QCD running coupling, so it is reasonable to
expect small-coupling methods to be applicable.
Nevertheless, the success of those methods may require a careful
account of effects beyond what is incorporated into the most
straightforward and familiar applications of collinear QCD
factorization.

Over the past three decades there has been significant progress in
extracting quantitative information about the partonic structure of
the nucleon from high-energy cross sections within the framework of
collinear factorization.  Indeed, a wealth of data from a wide range
of high-energy processes, covering many orders of magnitude of the
momentum transfer $Q$ and the Bjorken scaling variable $\xbj$,
can be described in terms of universal sets of parton distribution
functions (PDFs), both spin-averaged and spin-dependent---see
Refs.~\cite{Jimenez-Delgado:2013sma, Forte:2013wc, Blumlein:2012bf}
for recent reviews.
The essential elements of the collinear factorization framework
can be summarized as follows:
\begin{enumerate}

\item {\it Factorized formula.}
An observable, such as a structure function, $F$, is a convolution
integral over a longitudinal parton momentum fraction, $\xi$,
of a (hard) partonic coefficient function, $\widehat{H}$,
and a (soft) PDF, $f$,
\begin{equation}
F(\xbj, Q)
=  \int_{\xbj}^1 \frac{\diff{\xi}}{\xi}\,
   \widehat{H}\left(\frac{\xbj}{\xi},\frac{\mu}{Q}\right) f(\xi,\mu)\
+\ \order{\frac{m}{Q}},
\label{e.basicfact}
\end{equation}
where $Q$ is the hard scale and $\mu$ is a renormalization scale.
Here, and throughout this paper, $m$ will represent a generic mass
scale on the order of a hadron mass.  When different flavors of
partons are present, the convolution in addition involves matrix
multiplication.

\item {\it Longitudinal momentum.}
For collinear factorization, the convolution should only be over a
longitudinal momentum fraction.  The collinear approximations apply to
the limit that quantities such as intrinsic transverse momentum or
parton virtuality are $\order{m}$ and appear only in the power
suppressed error term typically as $\order{m^2/Q^2}$.

\item {\it Universal parton densities.}
The PDF $f(\xi, \mu)$ has a well-defined operator definition that
appears in a diverse class of collinear factorizable processes,
and so can be said to be universal.  The universality property
is especially central to global PDF
analyses~\cite{Jimenez-Delgado:2013sma, Forte:2013wc}.

\end{enumerate}

While the collinear factorization paradigm has been extremely useful
in applications at high energies, it is important to examine
the extent to which it can be practically utilized at the lower
range of energies of interest to studies of hadron structure in QCD,
where $\alpha_s$ may be small, but where effects from beyond the
usual kinematical collinear approximations become important.
Such effects include target mass corrections (TMC), higher twist
contributions, or intrinsic $k_T$ and parton virtuality.
Strictly speaking, collinear factorization derivations only apply
to the limit of small $m/Q$.  Nevertheless, $\alpha_s(Q)/\pi$
remains reasonably small even for values of $Q$ comparable to the
nucleon mass.  For example, $\tau$-lepton decays with $Q = 1.78$~GeV
are used in global extractions of the strong coupling, and find
$\alpha_s/\pi \approx 0.1$.~\cite{Bethke:2015etp}.

In the case of DIS, processes at scales of a few~GeV involve an
interesting mixture of perturbative and nonperturbative behavior.
For example, some consequences of a small coupling associated with
asymptotic freedom, such as approximate $Q^2$ scaling, persist
even at scales low enough for nonperturbative features like
resonances to be clearly observable (this is sometimes referred
to as ``precocious scaling'') \cite{Duke:1979na, Devoto:1983sh}. 
The observation of scalinglike behavior in certain observables in
kinematic regions where hadronic (resonance) degrees of freedom are
still prominent is related to the phenomenon of ``quark-hadron
duality,'' which characterizes the similarity between low-energy cross
sections, averaged over appropriate energy intervals, and those
computed from quarks and gluon in perturbative QCD \cite{Bloom:1971ye,
DeRujula:1976baf, Poggio:1975af, Ji:1994br}. Unraveling the dynamical
origin of this behavior remains a challenge for strong interaction
physics, and has motivated studies of the nature of the transition
from the perturbative to nonperturbative regimes of QCD (for a review
see Ref.~\cite{Melnitchouk:2005zr}).  Structure functions in the large-$\xbj$ region 
have also been used to explore the behavior of $\alpha_s(Q)$ in the nonperturbative limit~\cite{Courtoy:2013qca}.

Many techniques have been put forward for extending the basic
collinear factorization framework to accommodate quantitative
analyses of data at lower energy or larger $\xbj$.  Most aim to
accommodate small corrections from beyond strict collinearity.
One strategy has been to include certain classes of the $\order{m/Q}$
corrections in \eref{basicfact} by arguing that some types of
power-suppressed corrections are more important than others.
Another has been to perform all-order resummations of terms that
involve factors of $\ln(1-\xbj)$~\cite{Sterman:1987aj, Catani1989,
Dokshitzer:2005bf, Manohar:2003vb}.  In some approaches,
higher-twist operators in an operator product expansion (OPE) have
been able to be kept explicitly~\cite{Ellis:1982cd, Jaffe:1982pm}.

Of the various types of $1/Q$ power corrections, TMCs receive
particular attention in moderate- to low-$Q$ applications, where
$M/Q$-suppressed effects that are ordinarily neglected in standard
collinear factorization become important~\cite{Schienbein:2007gr}.
The most common approach to quantifying TMCs is based on the
pioneering work of Georgi and Politzer \cite{Georgi:1976ve}
and Nachtmann \cite{Nachtmann:1973mr}. 
It re-examines the OPE \cite{Wilson:1969zs, Brandt:1970kg,
Christ:1972ms} and includes some terms that would usually be
marked as power-suppressed, but neglects others such as
those associated with quark off-shellness.
This framework has been used to evaluate the TMCs for both
the spin-averaged \cite{Georgi:1976ve} and spin-dependent
\cite{Wandzura:1977qf} structure functions, at twist-two
and twist-three levels \cite{Blumlein:1998nv}. 
Corrections obtained in this way are often called
``kinematical higher twists,'' to distinguish them from
$1/Q$-suppressed ``dynamical higher twists'' that are
associated with multiparton operators in the OPE.

Strictly speaking, it is of course not possible to uniquely
decouple all TMCs from dynamical power corrections.
This was appreciated already in the early TMC work within the
OPE~\cite{Georgi:1976ve, DeRujula:1976ih, Gross:1976xt}, in the
context of the so-called ``threshold problem,'' whereby the target
mass corrected structure functions remain nonzero at $x=1$
\cite{Schienbein:2007gr, Bitar:1978cj, Johnson:1979ty, Steffens:2012jx}.
Later work~\cite{Ellis:1982cd} within a diagrammatic, momentum-space
approach extended the collinear factorization framework to lower $Q$
by accounting for multiparton correlations and TMCs up to
$\order{1/Q^2}$, including the effects of the parton transverse
momentum, $\kT$.
That analysis elucidated the relationship between the parton $\kT$
and the parton virtuality, and established a correspondence with
the earlier OPE formulation.

Most methods for dealing with target masses are rooted in a
fundamentally collinear picture, in that all nonperturbative
correlation functions depend only on collinear momentum fractions,
with an implicit assumption that corrections to purely collinear
kinematics are expressible as a series of powers in $m/Q$ or
$\alpha_s(Q)$, or both.
For moderately low $Q$, an alternative possibility is that a hard
factor can indeed be identified and expanded in small $\alpha_s(Q)$,
but that the associated nonperturbative factors become fundamentally
non-collinear.  In that case, multiple components of intrinsic
nonperturbative parton momentum might need to be included from the
outset, not merely in the form of small corrections to collinearity.
Parton correlation functions that go beyond the standard inclusive
collinear PDFs have a long history, and include objects like
transverse momentum dependent (TMD) parton distributions,
which include sensitivity to intrinsic transverse components
of parton momentum in addition to the usual longitudinal ones.
TMD PDFs are usually used for describing observables, such as in
semi-inclusive DIS, that  have direct sensitivity to intrinsic
parton $\kT$.  However, the particular kinematical scenarios at
moderate $Q$ or larger $\xbj$ might require similar shifts in the
underlying partonic picture, even at the totally inclusive level.

A complication with questions about the limitations of any one
approach, or about the advantages of one approach over another, is
that it is difficult to precisely estimate the sizes of errors without
greater knowledge of nonperturbative QCD than is currently available.
Nevertheless, improved methods for estimating the sizes of corrections
to factorization theorems are becoming more urgently needed for
addressing fundamental theoretical QCD questions in the relatively
complicated environment of moderate- to low-$Q$ physics.
A hope is that new efforts to understand PDFs from the lattice QCD
perspective may help.

The strategy of this paper is based on the observation that most
methods for deriving collinear factorization, such as the OPE
\cite{Wilson:1969zs, Brandt:1970kg, Christ:1972ms}, Libby-Sterman
style analyses of mass singularities \cite{Libby:1978qf}, or
soft-collinear effective theories~\cite{Becher:2014oda}, apply
generally to most simple renormalizable quantum field theories.
If a factorization formula is well-behaved in the context of QCD,
with all its complications from non-Abelian gauge invariance and
confinement, then it should certainly be well-behaved in a much
simpler renormalizable field theory without gauge degrees of freedom.
We will exploit this by exploring the limitations of factorization
derivations in a simple field theory of a quark coupling to a
scalar ``diquark'' to form a ``nucleon.''  We will use this to
stress test the standard collinear parton model kinematical
approximations.

We will argue, on the basis of the scalar diquark theory, that target
masses, quark masses, quark transverse momentum, and quark virtuality
are all likely to have similar quantitative importance at momentum
scales of order a few~GeV.
Moreover, the analysis will allow us to propose a factorization-based
notion of purely kinematical TMCs.
For the lowest $Q$ and largest $\xbj$ that typically define the
boundary of the DIS region, we find that corrections to a collinear
picture are not negligible, and new factorization theorems, with
correlation functions that depend on multiple components of
parton momentum, may be necessary.
Finally, we will illustrate the general usefulness of the scalar
diquark theory (or similar models) as a testing ground for the
approximations in a factorization derivation.
A factorization derivation deals, in essence, directly with
a power series expansion of the cross section in $m/Q$;
a factorization theorem is a characterization of the leading power.
Factorization is therefore the appropriate context for
characterizing the size and general behavior of power corrections.

This paper is organized as follows.
In \sref{model} we define the scalar diquark theory and
discuss its analogy with the pertinent features of QCD.
After providing the standard definition of inclusive DIS,
the full calculation with exact kinematics is presented
in \sref{exact}.  The computation includes all diagrams,
to lowest order in the coupling, that are necessary to
maintain electromagnetic gauge invariance.
We derive non-factorized expressions for the contributions
to the $F_1$ and $F_2$ structure functions from the ``handbag''
topology and $1/Q$-suppressed ``cat's ears'' diagrams.
The standard collinear factorization algorithm is presented
in \sref{factapp}, and the basic steps in the derivation
of the collinear PDF are outlined.
The results are found to be identical to those of the exact
calculation in the $m/Q \to \infty$ limit, but as $Q$ is lowered
one is able to study effects from nonvanishing $m/Q$ directly.
In \sref{pheno} we study these differences numerically,
with the goal of analyzing the relative importance of
different types of power corrections at moderate $Q$,
and identifying the regions of kinematics where the
collinearly factorized results may provide good approximations
to the exact structure functions.
Finally, in \sref{conclusions} we summarize our findings
and discuss their implications for future analyses.

%%%%%%%%%%%%%%%%%%%%%%%%%%%%%%%%%%%%%%%%%%%%%%%%%%%%%%%%%%%%%%%%%%%%%%%%
\section{DIS in a simple model}
\label{s.model}

%======================================================================%
\subsection{Definition}

We begin by describing the field theory we will use as a proxy for QCD
to highlight the salient aspects of factorization approximations at
moderate values of $Q$. Our results mainly concern the kinematics of
the process, and complications from the non-Abelian nature of the
full QCD theory do not directly affect the general conclusions.
The simplified theory is still sufficiently nontrivial that the
usual hurdles to deriving factorization in a renormalizable quantum
field theory are present.

The theory describes the interaction between a spin-1/2 ``nucleon''
with mass $\mhad$ represented by the field $\Psi_N$, a spin-1/2
``quark'' field $\psi_q$ with mass $\mquark$, and a scalar ``diquark''
state $\phi$ with mass $\mgluon$ that does not couple to the photon
but remains a spectator to the hard scattering from the quark.
The interaction Lagrangian density for this theory is given~by
a Yukawa-like interaction,
\begin{align}
\mathcal{L}_{\rm int}
  = -\lambda\, \overline{\Psi}_N\, \psi_q\, \phi\ +\ {\rm H. c.},
\label{e.lagrangian}
\end{align}
where the coupling $\lambda$ gives the strength of the
nucleon--quark--diquark interaction.
In this theory, the electron couples to quarks via electroweak
gauge bosons as in the standard model.  Furthermore, the theory is
renormalizable, and the basic derivation of factorization theorems
apply equally well to scattering processes here as to processes in QCD,
where non-Abelian gauge invariance leads to complications that make
factorization derivations more involved.
In practice, factorization means that $\order{Q}$ physics factorizes
from effects sensitive to intrinsic mass scales.
The simplified theory is ideal for stress-testing factorization
techniques generally before applying them to the more challenging
environment of a non-Abelian gauge theory such as QCD.

%======================================================================%
\subsection{Analogy with QCD}
\label{s.analogy}

%......................................................................%
\begin{figure}
\centering
\begin{tabular}{c@{\hspace*{1cm}}c@{\hspace*{1cm}}c}
\hspace*{-0.3cm}
\includegraphics[width=0.29\textwidth]{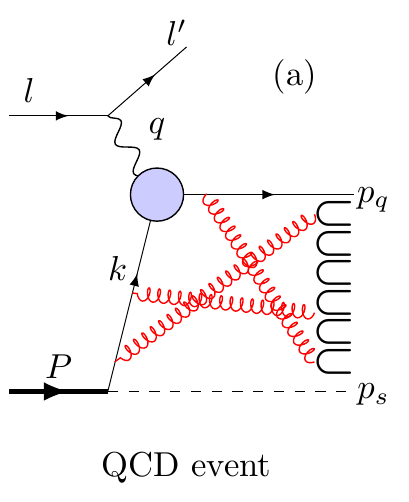}\ \
&
\includegraphics[width=0.29\textwidth]{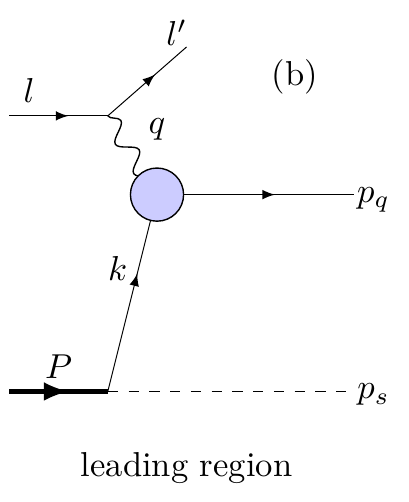}\ \
&
\includegraphics[width=0.29\textwidth]{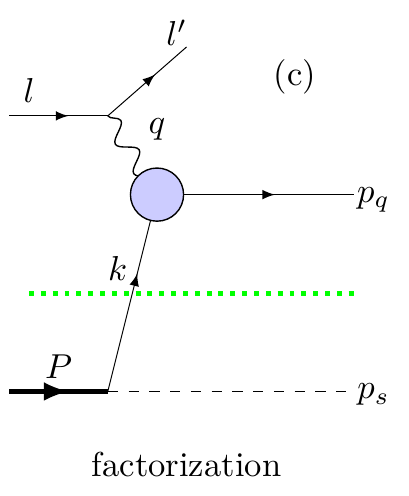}
\end{tabular}
\caption{
  The sequence of approximations leading to the canonical parton model
  picture:
  (a)~A~physical picture of the complete QCD event.
  The symbols $\subset$ represent the final state hadronization process.
  (b)~The leading-power topological region contributing to the inclusive
  cross section.
  (c)~The kinematical approximation (represented by the green dotted
  horizontal line) that produces the parton model cross section.
  The line is an instruction to replace the parton momentum by its
  approximated values (see \sref{factapp}).
  The momentum labels are discussed in the text.
}
\label{f.physicalpicture}
\end{figure}
%......................................................................%

The model described above is useful only to the extent that it
highlights important aspects of actual QCD interactions.
This is not a trivial point, since the handbag topology, while a
useful starting point, does not strictly capture the true nature
of QCD in DIS; a more accurate picture is probably closer to
Monte Carlo event generators.
Namely, partons generate showers of radiation both before and after
the collision, and an arrangement of final state partons undergoes
nonperturbative interactions to form a complex array of observable
hadrons.  This is illustrated in \fref{physicalpicture}(a).
This diagram emphasizes the physical picture of DIS:
a sea of parton fluctuations involving quarks, antiquarks and
gluons populates the rapidity interval between the incoming
hadron and struck quark rapidities, with the partons interacting
nonperturbatively to produce the final state hadrons.
[Final state gluons are not shown explicitly in
\fref{physicalpicture}(a).]

The factorization theorem for inclusive scattering states, in part,
that the sum of such diagrams may be approximated by the handbag
topology of \fref{physicalpicture}(b) in the limit of large $Q$.
The diagram in \fref{physicalpicture}(b) belongs to the leading
region for inclusive DIS.  Finally, a factorization formula emerges
once approximations are applied to the active parton momentum,
above and below the horizontal line in \fref{physicalpicture}(c)
separating the hard and soft parts of the diagram
(see Ref.~\cite{Collins:2011qcdbook} for more details).

The replacements in \fref{physicalpicture}, from (a) to (b)
and then (b) to (c), are only valid after integration over final
states that results in a cascade of cancellations of non-factorizing
effects.  The approximations therefore rely on the cross section
being fully inclusive.  Any map from exact underlying quark and gluon
degrees of freedom to the handbag picture is unavoidably indirect.
Nevertheless, for the factorization theorem to hold, it is a
\emph{necessary} condition that the approximations on parton
momentum represented by the horizontal line in
\fref{physicalpicture}(c) be at least roughly accurate.
Thus, the transition from (b) to (c) will be the focus of this
paper.  The main effect of that approximation is simply to alter
the kinematics of the handbag diagram.  We stress that such
approximations are at the core of QCD factorization theorems which
can also be studied in the context of the quark-diquark field theory.
We will review those approximations in Sec.~\ref{s.factapp}.

In our simple toy field theory, the magnitude of the factorization
error is fixed by the sizes of $m_q$ and $m_s$ relative to $Q$.
The same will be true in QCD, for the analogous quantities.
These parameters determine the size of the small components of
parton four-momentum related to $k^2$ and $\Tsc{k}{}$.  Other aspects of the
quark-diquark theory, such as the dominant $\Tsc{k}{}$ power-law of correlation functions
at large $\Tsc{k}{}$, are also the same in QCD.  The main difference
between QCD and the toy theory is that, while the values of $\mquark$
and $\mgluon$ are exactly fixed by the Lagrangian (and by our
restriction to the lowest-order graph) in the diquark theory,
in QCD the effective parton and spectator masses generally have
a spectrum of values that depend on $\xbj$, $\Tsc{k}{}$ and $Q$
and intrinsic properties of the nucleon wave function.
The kinematically allowed phase space grows with decreasing $\xbj$
and increasing $Q$, accommodating more of the soft radiation
sketched in \fref{physicalpicture}(a).  Thus, the scales
analogous to $\mquark$ and $\mgluon$ will generally acquire
nontrivial $\xbj$ and $Q$ dependence in QCD.

In both theories, however, $|k^2|$ and $\Tscsq{k}{}$ need to be small
relative to $Q^2$ to give the $m/Q$ suppression of neglected terms
that is necessary for the factorization theorem in \eref{basicfact}
to hold.  If $\mquark$ and $\mgluon$ are fixed to reasonable values
for a given range of kinematics, and if the integration over $\Tsc{k}{}$
is dominated by $\Tsc{k}{} \ll Q$, then we may verify directly that
the parton model approximations are good for the quark-diquark theory.
Showing this directly lends some support to the same approximations in QCD.
Conversely, if the approximations fail dramatically in the toy theory,
then it is unlikely that they are safe in QCD for the same kinematical
region, particularly given the additional complications with
non-Abelian gauge invariance, strong coupling, and nonperturbative
hadronization.

Carrying this out requires a reasonable set of estimates for
$\mgluon$ and $\mquark$ for a specified ranges of kinematics.
For $Q\sim$~several GeV, the requirement that $m/Q$ is small
implies that $\mquark$ should be no larger than several hundred~MeV
and $\mgluon$ should be such that $|k^2|$ is also no larger than
several hundred MeV for small $\Tsc{k}{}$.  Unfortunately, there
are, to our knowledge, no systematic methods for precisely
estimating values for the small components of parton momentum
like $\mquark$ and $|k^2|$. 
On the other hand, phenomenological studies of transverse
momentum dependence in semi-inclusive DIS suggest typical
ranges for these parameters.  
Extractions of TMD functions find typical magnitudes for the
intrinsic transverse momentum width between $\approx500$~MeV and
800~MeV~\cite{Feynman:1978dt, Anselmino:2013lza, Signori:2013mda}. 
Since $\mquark$ and $\mgluon$ determine the widths and shapes of
the $\Tsc{k}{}$ distribution, these estimates provide reasonable
lower bounds on $\mquark$ and $\mgluon$.
Earlier estimates gave smaller values. For example, a value of $\langle \Tsc{k}{} \rangle \sim 300$~MeV is roughly 
consistent with both the zero point energy of bag models as well 
as non-relativistic constituent quark models~\cite{Bhaduri:1988gc}, and 
this is the value quoted in Ref.~\cite{Georgi:1976ve}. It is interesting to ask why phenomenological 
extractions tend to produce broader nonperturbative distributions than these expectations. 
(See also the discussion in Ref.~\cite{Feynman:1978dt}.) For now we leave this to be addressed in future work.

In this analysis we will use a range of values for $\mquark$ and
$\mgluon$ motivated by the above estimates, and examine the sensitivity
to their variation for $Q \sim$~1--2~GeV and moderate $\xbj$.
Sensitivity to the exact values of these parameters will be
interpreted as a sign that extra care may be needed when
estimating their effects on power corrections.
We will return to the question of exact values for $m_q$ and $m_s$
in \sref{paramvals}, after examining DIS kinematics in more detail.

%======================================================================%
\subsection{Structure tensors}

Let us review the standard notation of the inclusive DIS process
$e(\ell) + N(P) \to e(\ell') + X(p_X)$
in \fref{physicalpicture}, where $\ell$ and $\ell'$ are the
initial and final lepton four-momenta, $P$ is the four-momentum
of the nucleon, and $p_X = \jet+\spectator$ is the four-momentum
of the inclusive hadronic state $X$.
It will be convenient for our analysis to work in the Breit frame,
where the nucleon moves along the $+z$ direction and the virtual
photon moves along the $-z$ axis with zero energy.
We will use light-front coordinates, in which a four-vector
$v^\mu = (v^+, v^-, \T{v}{})$ has ``$\pm$'' components
$v^\pm = (v^0 \pm v^z)/\sqrt2$ and transverse component $\T{v}{}$.
The four-momenta of the nucleon and the exchanged photon
($q = \ell - \ell'$) can then be written as
\begin{align}
P^\mu &= 
  \parz{ \frac{Q}{\xn \sqrt{2}},
		     \frac{\xn \mhad^2}{Q \sqrt{2} },
	 	     \T{0}{}},			
\label{e.P}\\
q^\mu &= 
  \parz{-\frac{Q}{\sqrt{2}},
	 	     \frac{Q}{\sqrt{2}},
		     \T{0}{}},
\label{e.q}
\end{align}
where $Q \equiv \sqrt{-q^2}$ is the magnitude of the four-momentum
transfer, and
\begin{align}
\xbj & \equiv \frac{Q^2}{2 P \cdot q},
\label{e.xbj}\\
\xn  & \equiv -\frac{q^+}{P^+}
 = \frac{2 \xbj}{1 + \sqrt{1 + 4 \xbj^2 \mhad^2/Q^2}}
\label{e.nacx}  
\end{align}
are the Bjorken and Nachtmann scaling variables, respectively.
The Bjorken variable $\xbj$ can also be written in terms of the
Nachtmann variable,
\begin{align}
\xbj = \frac{\xn}{(1 - \xn^2 \mhad^2/Q^2)}.
\end{align}
Considering the leading region, \fref{physicalpicture}(b),
the final state quark (or ``jet'') momentum is $\jet$,
and the momentum of the spectator system is $\spectator$, with
\begin{equation}
\jet^2 = \mquark^2 \, , \qquad \spectator^2 = \mgluon^2 \, .
\end{equation}
We also define a momentum transfer variable,
\begin{equation}
k \equiv \jet - q = P - \spectator \, .
\end{equation}
In a handbag diagram [see \fref{basicmodel}(a) below],
$k$ would be the momentum of the incoming struck quark.
The invariant mass squared of the photon--nucleon system is
\begin{align}
W^2 &= (P + q)^2
     = (\jet + \spectator)^2
     = M^2 + \frac{Q^2 (1-\xbj)}{\xbj}.
\label{e.W}
\end{align}

The boost-invariant cross section for the inclusive DIS process is
\begin{equation}
E' \frac{\diff{\sigma}{}}{\diff{^3 \3{\ell}' }{}}
= \frac{\alfa^2}{2 \pi (s - \mhad^2) Q^4} L_{\mu\nu} W^{\mu\nu},
\label{e.cs1}
\end{equation}
where $\alfa$ is the electromagnetic fine structure constant, $E'$ is the final 
lepton energy, and $s$ is the usual Mandelstam variable. The only approximation 
is to neglect the lepton mass in the flux factor.
The leptonic tensor is
$L_{\mu\nu}
 = 2(\ell_\mu \ell'_\nu + \ell'_\mu \ell_\nu - g_{\mu\nu} \ell\cdot{\ell'})$
is the leptonic tensor. 
We are most interested in the hadronic tensor,
\begin{equation}
W^{\mu\nu}(P,q) = \sum_X  
 \langle P, S | j^\mu(0) | X \rangle  \langle X | j^\nu(0) | P, S \rangle \, 
(2 \pi)^4 \delta^{(4)} (P + q - p_X) . \label{e.hadtens}
\end{equation}
Here $\sum_X$ represents the inclusive integration over all hadronic final states 
with overall four-momentum $p_X$. Note that all factors of $\alpha$ appear 
in the prefactor in \eref{cs1}. Also, we have moved a conventional $1/(4 \pi)$ 
factor from the definition of the hadronic tensor into the overall factor in 
\eref{cs1} to minimize the number of factors of $\pi$ that need to be accounted for 
in intermediate steps. 
For the scattering of an unpolarized lepton from an unpolarized
nucleon, the hadronic tensor $W^{\mu\nu}$ is usually expressed in
terms of the spin-averaged structure functions $F_1$ and $F_2$,
\begin{align}
W^{\mu \nu}(P,q) 
&= \parz{-g^{\mu \nu} + \frac{q^\mu q^\nu}{q^2} } 
   F_1 \parz{\xn,Q^2}					\notag \\
&+ \parz{P^\mu - \frac{P \cdot q}{q^2} q^\mu}
   \parz{P^\nu - \frac{P \cdot q}{q^2} q^\nu}
   \frac{F_2\parz{\xn, Q^2}}{P \cdot q}.
\label{e.Wmunu}
\end{align}
The structure functions are obtained from the hadronic tensor
by applying projection operators,
\begin{align}
F_i\parz{\xn,Q^2} = {\rm P}_i^{\mu\nu}\, W_{\mu\nu}(P,q),\ \ \ \ \
i=1,2,
\end{align}
where
\begin{subequations}
\begin{align}
{\rm P}_1^{\mu\nu} 
&= -\frac{1}{2} {\rm P}_g^{\mu\nu}
   +\frac{2 Q^2 \xn^2}{(\mhad^2 \xn^2 + Q^2)^2} {\rm P}_{PP}^{\mu\nu},
\label{e.P1}\\
{\rm P}_2^{\mu\nu}
&= \frac{12 Q^4 \xn^3 \left(Q^2-\mhad^2 \xn^2\right)}
	 {\left(Q^2 + \mhad^2 \xn^2\right)^4}
    \left( {\rm P}_{PP}^{\mu\nu}
	 -\frac{\left(\mhad^2 \xn^2+Q^2\right)^2}{12 Q^2 \xn^2}
	   {\rm P}_g^{\mu\nu}\right),
\label{e.P2}
\end{align}
\label{e.F12proj}
\end{subequations}
with the components 
\begin{align}
{\rm P}_g^{\mu\nu} = g^{\mu\nu},\;\quad
{\rm P}_{PP}^{\mu\nu} = P^\mu P^\nu. 
\label{e.PgPP}
\end{align}
See Ref.~\cite{Bacchetta:2006tn} for a full structure decomposition 
of SIDIS with spin and azimuthal dependence and exact kinematics.

In \eref{Wmunu} we have written the structure functions in terms of
the Nachtmann $\xn$ variable instead of Bjorken $\xbj$, as is more
commonly presented in the literature.  The reason is that $\xn$ is the
natural scaling variable in the parton model approximation $k^+
\approx -q^+$ when $\mhad$ is not set to zero.  In the limit that
power suppressed terms can be dropped, the two scaling variables are
equal,
\begin{align}
\xn = \xbj + \order{\frac{\xbj^2 \mhad^2}{Q^2}},
\end{align}
although we stress that the $\xn \approx \xbj$ approximation is not
generally necessary and is separate from the approximations needed to
factorize short- and long-distance physics in a theory with
interactions.  Both $\{ \xn, Q \}$ and $\{ \xbj, Q \}$ are equally
valid as independent kinematic variables; since $\xn$ is the natural
variable when hadron masses are not neglected, we will use it
everywhere unless specified otherwise.

%%%%%%%%%%%%%%%%%%%%%%%%%%%%%%%%%%%%%%%%%%%%%%%%%%%%%%%%%%%%%%%%%%%%%%%%
\section{Exact kinematics}
\label{s.exact}

Having defined the model and the quantities of interest,
in this section we calculate the DIS structure functions from the
Lagrangian $\mathcal{L}_{\rm int}$ in \eref{lagrangian} at the
lowest nontrivial order, $\order{\alfa \lambda^2}$.
The corresponding graphs derived from $\mathcal{L}_{\rm int}$
are shown in \fref{basicmodel}.
Graph~(A) has the familiar handbag diagram topology, while
graphs (B) and (C) are power-suppressed at large $Q$
but are needed for exact electromagnetic gauge invariance---see Appendix~\ref{a.EMGaugeInv}.
We exclude the elastic limit of $\xbj = 1$ and require
strictly $W > M$, so that diagrams with an on-shell nucleon
in the final state are forbidden.

Graphs (B) and (C) represent the direct coupling of the photon
to the nucleon, with production of a far off-shell nucleon in
the intermediate state.  In the quark-diquark field theory the
coupling is point-like, while in QCD it corresponds to a
higher-twist interaction internal to the nucleon wave function,
with the final state quark interacting with the nucleon remnant
to form a highly virtual intermediate state.

%......................................................................%
\begin{figure}
\hspace*{-0.5cm}
\includegraphics[width=0.31\textwidth]{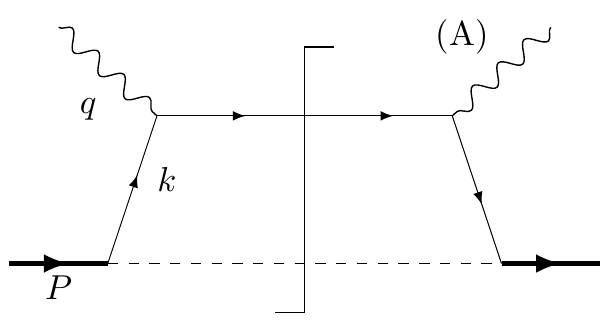}\ \ \ \
\includegraphics[width=0.31\textwidth]{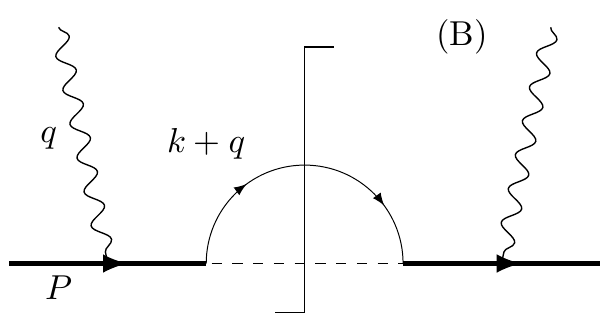}\ \ \ \
\includegraphics[width=0.31\textwidth]{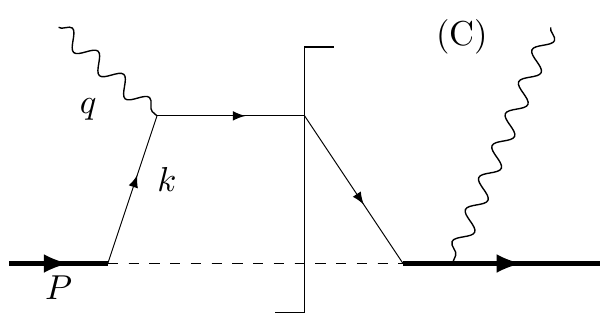}
\caption{
  Contributions to the hadronic tensor from diagrams allowed
  by the interaction Lagrangian~(\ref{e.lagrangian}) to
  $\order{\alpha\lambda^2}$
  in the couplings.
  Graph~(A) is a manifestation of the familiar handbag diagram
  and represents the topology of the leading region.
  Graphs~(B) and (C) are suppressed by powers of $1/Q$ when
  $\Tsc{k}{}$ is small, but are needed for gauge invariance.
  The Hermitian conjugate for (C) is not shown.
  The momenta on the various legs are as indicated.
}
\label{f.basicmodel}
\end{figure}
%......................................................................%

We begin by presenting the organization of the calculation of
the graphs in \fref{basicmodel}, with no approximations whatsoever
on kinematics.  Of course, the result will not be factorized.
Later, we will compare with the canonical parton model approximations
that factorize the graphs into a hard collision and a PDF contribution.

The exact calculation is organized by separating the integrand of
the hadronic tensor into factors representing different parts of
the squared amplitude,
\begin{align}
W^{\mu\nu}(P,q)
&{}= 
\sum_{j \in {\rm \; graphs}}\frac{1}{2} \int
  \frac{\diff{k^+}{} \diff{k^-}{} \diff{^2 \T{k}{} }{} }
       {(2 \pi)^2}\,
  \jac{}\, \num_j^{\mu \nu}\, \prop_j\,
  \delta(k^- - k^-_{\rm sol})\, \delta (k^+ - k^+_{\rm sol}),
\label{e.unappintexact} 
\end{align}
where $k$ is the four-momentum of the interacting parton,
and the sum over $j$ runs over the graphs labeled by
$j \in \{ A, B, C \}$.
The propagator denominators in \eref{unappintexact} have been gathered
into the factor $\prop_j$, and the traces over the $\gamma$ matrices
are denoted by $\num_j^{\mu \nu}$.  The resulting Jacobian factor
associated with the integration over $k^\pm$ is denoted as $\jac{}$. To simplify notation, we will fix $\lambda=\sqrt{2}$ and drop all explicit factors of $\lambda^2$ throughout the rest of this article.
The $\delta$-functions stem from the on-shell conditions for the final
state quark and scalar diquark,
\begin{subequations}
\begin{align}
(q + k)^2 - \mquark^2 &{}= 0\, , \\
(P - k)^2 - \mgluon^2 &{}= 0\, .
\end{align}
\end{subequations}
Solving this system of equations for $k^+ \equiv \xi P^+$ and $k^-$
gives two solutions for $k^-$. In the limit of $Q \to \infty$ with
$\xn$ and $\kT$ fixed, the two solutions behave as $k^- \sim \infty$
and $k^- \sim 0$, respectively.  Selecting the latter as the physically
relevant solution for DIS, we obtain the values of the light-cone
parton momenta $k^\pm_{\rm sol}$ with on-shell final state quark
and diquark,
\begin{subequations}
\label{e.kplusminus}
\begin{align}
k^- =\, k^-_{\rm sol}
&\equiv \frac{ \sqrt{\Delta} - Q^2 (1 - \xn)
	     - \xn \big( m_s^2 - m_q^2 - M^2 (1 - \xn) \big)}
	     {2 \sqrt{2}~Q~(1 - \xn)},
\label{e.kplus}\\
k^+ =\, k^+_{\rm sol}
&\equiv \frac{\kT^2 + m_q^2 + Q (Q + \sqrt{2} k^-)}
	{\sqrt{2} (Q + \sqrt{2} k^-)},
\label{e.kminus}
\end{align}
\end{subequations}
where $k_T^2=\Tscsqb{k}{}$, and the discriminant $\Delta$ is
\begin{align}
\Delta =& \big[ Q^2 (1-\xn) 
               - \xn \left( M^2 (1-\xn) + m_q^2 - m_s^2 \right)
           \big]^2\notag\\  
        & -4 \xn ( 1-\xn)
          \big[ \kT^2 (Q^2 + \xn M^2) 
                - Q^2 M^2 (1-\xn)
	              + Q^2 m_s^2 + \xn M^2 m_q^2
          \big].
\label{e.Delta}
\end{align}
The parton virtuality is obtained by substituting~\erefs{kplus}{kminus} into
\begin{equation}
k^2 = 2 k^+ k^- - \Tscsq{k}{} \, . 
\end{equation}
The Jacobian factor in \eref{unappintexact} is 
\begin{align}
\jac{}
= \frac{\xn Q~(2k^-+\sqrt{2}Q)}
       {4 (1-\xn) k^- Q^2 (\sqrt{2} k^- + 2Q)
	+ 2\sqrt{2}
	  \big[Q^4 (1-\xn) - (\kT^2+m_q^2) \xn (Q^2 + \xn\, M^2)
	  \big]}.
\label{e.jac}
\end{align}
For this article, we are interested in the small-$\left|k^2\right|$ region where a parton model approximation might be reasonable. The $k^-$ solution corresponding to large $\left|k^2\right|$ is dealt with in an $\order{\lambda^2}$ treatment of the hard part. The exact propagator factors for each of the contributions in
\fref{basicmodel} are
\begin{subequations}
\label{e.prop}
\begin{align}					              
\prop_{\rm A}
&= \frac{1}{(k^2 - \mquark^2)^2},
\label{e.propA}		\\
\prop_{\rm B}
&= \frac{1}{\big( (P+q)^2 - \mhad^2 \big)^2}
 = \frac{\xn^2}{\big( Q^2 (1-\xn)- \mhad^2 \xn^2\big)^2},
\label{e.propB} 	\\
\prop_{\rm C}
&=\frac{1}{(k^2 - \mquark^2)}
  \frac{\xn}{\big( Q^2 (1-\xn)- \mhad^2 \xn^2\big)}.
\label{e.propC}
\end{align}
\end{subequations}
The numerator factors
$\num_j^{\mu\nu} = \num_j^{\mu\nu}(P,k,\mquark,\mgluon)$ are
obtained from the Dirac traces in each graph in \fref{basicmodel},
\begin{subequations}
\label{e.Tj}
\begin{align}
\num_{\rm A}^{\mu\nu}
&= {\rm Tr}
   \left[ 
    (\slashed{P} + M)
    (\slashed{k} + m_q) 
    \gamma^\mu
    (\slashed{k} +\slashed{q}+ m_q) 
    \gamma^\nu
    (\slashed{k} + m_q) 
   \right],					\\
\num_{\rm B}^{\mu\nu}
&= {\rm Tr}
   \left[ 
    (\slashed{P} + M)
    \gamma^\mu
    (\slashed{P} +\slashed{q}+ M) 
    (\slashed{k}+\slashed{q} + m_q) 
    (\slashed{P} +\slashed{q}+ M) 
    \gamma^\nu
   \right],					\\
\num_{\rm C}^{\mu\nu}
&= 2\, {\rm Tr}
   \left[ 
    (\slashed{P} + M)
    (\slashed{k} + m_q) 
    \gamma^\mu
    (\slashed{k} +\slashed{q}+ m_q) 
    (\slashed{P} +\slashed{q}+ M) 
    \gamma^\nu
   \right],
\end{align}
\end{subequations}
where the factor of $2$ in $\num_{\rm C}^{\mu\nu}$ accounts
for the Hermitian conjugate of \fref{basicmodel}(C).
In evaluating the traces \eref{Tj}, it will be convenient
to define the projected quantities
\begin{equation}
\num_j^g{} = {\rm P}_g^{\mu\nu}\, \num_{j\, \mu\nu}\, ,  \qquad
\num_j^{PP}{} = {\rm P}_{PP}^{\mu\nu}\, \num_{j\, \mu\nu}\, .
\label{e.num_proj}
\end{equation}
Evaluating the projections explicitly, 
\begin{subequations}
\label{e.TgPP}
\begin{align}
\num_{\rm A}^g
=& -8 \left[
    2 (P \cdot k + m_q M)\, k \cdot q
  + (k^2 - 3 m_q^2)\, P \cdot k
  - 2 M m_q^3 
  + (m_q^2-k^2)\, P \cdot q
  \vphantom{\frac{M}{M}}\right],	\\
\num_{\rm B}^g
=&\quad8 \left[
    2 M^3 m_q 
  + P \cdot k\, (2 M^2 - Q^2) 
  - 2 (M^2 + M m_q)\, Q^2 
  \vphantom{\frac{M}{M}}\right.		\notag\\
&  \left.\quad\quad
  + 2 k \cdot q\, (M^2 - P \cdot q)
  + [2 (M^2 + M m_q) + Q^2]\, P \cdot q
  \vphantom{\frac{M}{M}}\right],	\\
\num_{\rm C}^g
=& -16 \left[
  -2 (P \cdot k)^2 
  + k^2 M^2 
  + (M^2-m_q M)\, k \cdot q
  - M^2 m_q^2 
  + 2 M m_q Q^2 
  \vphantom{\frac{M}{M}}\right.		\notag\\
&  \left.\quad\quad
  + (m_q^2-M m_q) P \cdot q
  - 2 P \cdot k\, (k \cdot q + M m_q - Q^2 + P \cdot q)
  \vphantom{\frac{M}{M}}\right],	\\
\num_{\rm A}^{PP}
=&\quad 4 \left[ \vphantom{\frac{M}{M}} 
   4 (P \cdot k)^3 
  + 4 (P \cdot k)^2 (M m_q + P \cdot q)
  \vphantom{\frac{M}{M}}\right.		\notag\\
&  \left.
  \quad
  \quad
  - M\, P \cdot k\,
    (3 k^2 M + 2M\, k \cdot q - 3 M m_q^2 - 4 m_q\, P\cdot q) 
  \vphantom{\frac{M}{M}}\right.		\notag\\
&  \left.
  \quad
  \quad
  - M^3 m_q (k^2 + 2 k \cdot q - m_q^2) 
  - M^2 (k^2 - m_q^2)\, P \cdot q
  \vphantom{\frac{M}{M}} \right],	\\
\num_{\rm B}^{PP}
=&\quad 4 M^2 
  \left[ \vphantom{\frac{M}{M}} 
     P \cdot k\, (4 M^2 + Q^2) 
   + 4 M^2 (k \cdot q + M m_q) 
   - Q^2 (4 M^2 + M m_q)
  \vphantom{\frac{M}{M}}\right.		\notag\\
&  \left. \quad\quad
   + [2 k \cdot q + 4 (M^2 + M m_q) - Q^2]\, P \cdot q
  \vphantom{\frac{M}{M}} \right],	\\
\num_{\rm C}^{PP}
=&\quad 8 M 
  \left[ \vphantom{\frac{M}{M}} 
     4 M (P \cdot k)^2 
     + M\, P \cdot k\, (2 k \cdot q + 4 M m_q - Q^2)
  \vphantom{\frac{M}{M}}\right.		\notag\\
&  \left. \quad\quad
   - M^2 [2 M (k^2 + k \cdot q - m_q^2) + m_q Q^2]
  \vphantom{\frac{M}{M}}\right.		\notag\\
&  \left. \quad\quad
  - [k^2 M - (2 M + m_q) (2 P \cdot k + M m_q)]\, P \cdot q
  \vphantom{\frac{M}{M}} \right] \, .
\end{align} 
\end{subequations}
Putting all the components together, the exact nucleon structure
functions $F_{1,2}$ can be written in terms of the $\kT$-unintegrated
distributions,\footnote{Note that these are not PDFs, which are only
defined after factorizing approximations are applied.}
\begin{subequations}
\label{e.F12kTdef}
\begin{eqnarray}
F_1\parz{\xn,Q^2}
&=& \int\frac{ \diff{^2 \T{k}{} }{} }{(2\pi)^2}\,
    {\cal F}_1(\xn,Q^2,\kT^2),
\label{e.F1kTdef}					\\
F_2\parz{\xn,Q^2}
&=&  \int\frac{ \diff{^2 \T{k}{} }{} }{(2\pi)^2}\,
    2\xn\, {\cal F}_2(\xn,Q^2,\kT^2),
\label{e.F2kTdef}
\end{eqnarray}
\end{subequations}
where 
\begin{subequations}
\label{e.F12kT}
\begin{eqnarray}
{\cal F}_1\parz{\xn,Q^2,\kT^2}
&=& 
    \jac{} \sum_j
    \left(- \frac{1}{2} \num_j^{g}
	  + \frac{2 Q^2 \xn^2}{(\mhad^2 \xn^2 + Q^2)^2} \num_j^{PP}
    \right)
    \prop_j,
\label{e.F1kT}						\\
2\xn {\cal F}_2\parz{\xn,Q^2,\kT^2}
&=& 
    \frac{12 Q^4 \xn^3 (Q^2-\mhad^2 \xn^2)}
	 {(Q^2 + \mhad^2 \xn^2)^4}			\nonumber\\
& & \times
    \jac{} \sum_j
    \left( \num_j^{PP}
	 - \frac{(\mhad^2 \xn^2+Q^2)^2}{12 Q^2 \xn^2} \num_j^{g}
    \right)
    \prop_j.
\label{e.F2kT}
\end{eqnarray}
\end{subequations}
For later convenience, the function ${\cal F}_2$ in
Eqs.~(\ref{e.F2kTdef}) and (\ref{e.F2kT}) has been defined with
a factor $2\xn$ pulled out in order to more directly compare the
behavior of the $\kT$ dependence of the $\kT$-unintegrated functions
(see \sref{pheno} below). 

Note that exact kinematics impose a specific upper bound on $\Tsc{k}{}$.
To determine its value, write $W$ in the center-of-mass (c.m.) system,
\begin{equation}
W = \jet^0 + \spectator^0 \Big|_{\rm c.m.}
  = \sqrt{\mquark^2 + \Tscsq{k}{} + k_z^2} 
	 + \sqrt{\mgluon^2 + \Tscsq{k}{} + k_z^2}
    \Big|_{\rm c.m.}.
\end{equation}
For fixed external kinematics, the maximum $\Tsc{k}{}$ occurs
when $k_z = 0$.  Setting
\begin{equation}
\sqrt{\mquark^2 + \ktmaxsq} + \sqrt{\mgluon^2 + \ktmaxsq} = W \, 
\end{equation}
and solving for $\ktmax$ gives
\begin{align}
\ktmax
{}&= 
\sqrt{
  \frac{
    \big[ \xbj (\mhad^2 - (\mquark + \mgluon)^2) + Q^2 (1-\xbj) \big]
    \big[ \xbj \left( \mhad^2 - (\mquark-\mgluon)^2 \right)
         + Q^2 (1-\xbj)
    \big]
    }{4 \xbj \big[ Q^2 (1-\xbj) + \mhad^2 \xbj \big]}
} \, ,
\label{e.ktmax} 
\end{align}
where \eref{W} has been used for $W$.
Results for the exact structure functions will be shown in
Sec.~\ref{s.pheno}.

%%%%%%%%%%%%%%%%%%%%%%%%%%%%%%%%%%%%%%%%%%%%%%%%%%%%%%%%%%%%%%%%%%%%%%%%
\section{Factorization}
\label{s.factapp}

In this section, we review the minimal kinematic approximations
needed for standard factorization with low-order handbag graphs such
as in \fref{handbag}.  More details with extensive discussion of the
justification for the applicability of factorization may be found,
for example, in Sec.~6.1.1 of Ref.~\cite{Collins:2011qcdbook}.

The first step in a collinear factorization derivation in DIS is to
identify and restrict attention to leading (in $m/Q$) region graphical
topologies.  One such configuration, and the only one contributing
at zeroth order coupling in the hard part, is the handbag topology of
\fref{handbag}(a), with two final state jets:
  one with momentum $k'=k+q$,
  and the other with momentum $P-k$.
The ``cat's ears'' graph topologies, corresponding to
Figs.~\ref{f.basicmodel}(B) and \ref{f.basicmodel}(C), are suppressed
by powers of $1/Q^2$ and so do not contribute in the leading power
approximation.

%......................................................................%
\begin{figure}
\centering
\includegraphics[width=0.6\textwidth]{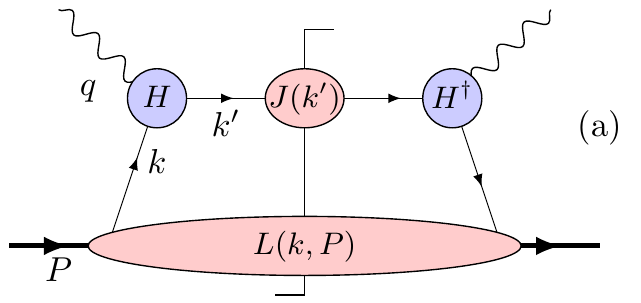}\\
\includegraphics[width=0.6\textwidth]{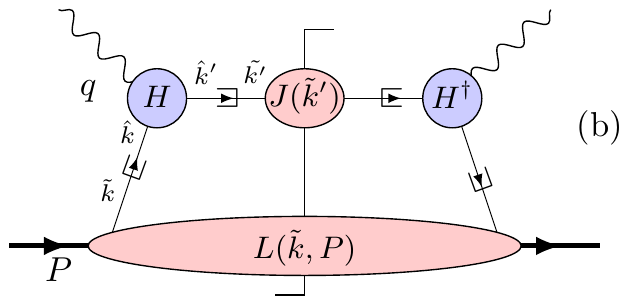}\\
\includegraphics[width=0.6\textwidth]{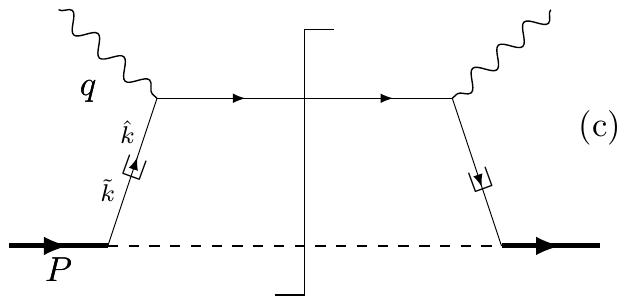}
\caption{
  The steps in the usual factorization approximation applied
  to a handbag topology.
  (a)~Unapproximated handbag topology, with $H$ denoting the
  hard scattering of a virtual photon from a quark with
  momentum $k$ to one with momentum $k'=k+q$,\
  $J(k')$ is the jet function, and
  $L(k,P)$ is the soft target amplitude.
  (b)~Handbag diagram with standard factorization, with the
  parton momentum approximated by $\hat{k}$ in the hard
  function $H$, and by $\tilde{k}$ in the jet and soft functions.
  The hooks represent the point of application of kinematic
  approximations on parton momentum.
  (c)~Application of the $\order{\lambda^2}$ contribution
  in the theory from \sref{model}.
}
\label{f.handbag}
\end{figure}
%......................................................................%

The contribution to the hadronic tensor from the amplitude in
\fref{handbag}(a) has the general form
\begin{equation}
W^{\mu\nu}(P,q) 
  = \int\!\frac{\diff{^4 k}{}}{(2 \pi)^4}\,{\rm Tr}
  \big[ H^\mu(k,k')\, J(k')\, H^{\nu\dagger}(k,k')\, L(k,P) \big].
\label{e.unappint}
\end{equation}
Here $H^\mu(k,k')$ and $H^{\nu\dagger}(k,k')$ represent the hard
scattering blobs in \fref{handbag}(a), where all internal lines
off-shell by at least $\order{Q^2}$.
The target, $L(k,P)$, and jet, $J(k')$, blobs have internal lines
off-shell by $\order{m^2}$, where the generic hadronic mass scale
$m \in \{ \mquark,\, \mgluon,\, \mhad \}$.
The parton lines that connect the various blobs have small
off-shellness, with $k^2$ and $k'^2 \sim \order{m^2}$.
In the Breit frame $k^+ \sim \order{Q}$.
The low transverse momentum region is where
    $\Tsc{k}{} \sim \order{\T{m}{}}$,
where $\T{m}{}$ denotes the transverse momentum components
of the parton momentum, each of which is of $\order{m}$.
The power counting for the struck parton momentum is therefore
\begin{align}
k & \sim \parz{\order{Q},\,
            \order{\frac{m^2}{Q}},\,
            \order{\T{m}{}}}.
\label{e.powercounting}
\end{align} 
We remind the reader that $m$ symbolizes any typical hadronic mass scale.
To factorize the cross section, one exploits Eq.~\eqref{e.powercounting}
to justify a standard set of kinematic approximations that we now review.

In the hard subgraphs, terms proportional to $k^2$ or $k'^2$ are small
relative to the $\order{Q^2}$ off-shellness of the propagators.
Since $k \cdot q = k^+ q^- + \order{m^2}$, the replacement of
$k \cdot q \to k^+ q^-$ in the hard blobs therefore introduces
only $\order{m^2/Q^2}$ suppressed errors at small $\Tsc{k}{}$.
Thus, the momenta in the hard parts are replaced by partonic
``hatted'' variables $\hat{k}$ and $\hat{k}'$,
\begin{subequations}
\label{e.khatall}
\begin{align}
k  & \to \hat{k}\  
      \equiv\ \parz{\hat{k}^+, 0, \T{0}{}} \, ; 
      \qquad \mhad^2/Q^2 \to 0 \, , 
\label{e.khat1}\\
k' &\to \hat{k}'\ =\ \hat{k} + q,
\end{align}
with
\begin{align}
\hat{k}^2 = \hat{k}'^2=0 \, . 
\label{e.khat2}
\end{align}
\end{subequations}
These equations give
\begin{subequations}
\label{e.hatk}
\begin{align}
\hat{k}^+ &=\, \xn \, P^+\ \stackrel{\xn \to \xbj}{=}\ \xbj P^+,
\label{e.kapp} \\
%
% \hat{k}'&= \parz{0, q^-, \T{0}{}}, \; (\xn \to \xbj) \, .
\hat{k}'\ &= \parz{0,\, q^-,\, \T{0}{}}.
\label{e.k2app}
\end{align}
\end{subequations}
The replacement $H^\mu(k,k') \to H^\mu(\hat{k},\hat{k}')$ is
therefore a good approximation up to $\order{m^2/Q^2}$ corrections. 
(The replacement of $\xn$ by $\xbj$ is not necessary to obtain
factorization, but it is conventional to use $\xbj$.)

Internal lines in the lower blob $L(k,P)$ are off-shell by
$\order{m^2}$.  With the replacement of
$k^+ \to \xbj P^+ + \order{m^2/Q}$, 
\begin{align}
k^2 
& = 2 k^+ k^- - \Tscsq{k}{} \notag\\
& = \underbrace{2 (\xbj P^+) k^- - \Tscsq{k}{}}_{\order{m^2}}\
  +\ \order{m^4/Q^2},					\\ 
(P + k)^2 
& = M^2 + 2 P^+ k^- + 2 P^- k^+ + 2 k^+ k^- - \Tscsq{k}{}	\notag\\
& = \underbrace{M^2 + 2 P^+ k^- + 2 P^- (\xbj P^+)
		+ 2 (\xbj P^+) k^- - \Tscsq{k}{}}_{\order{m^2}}\
  +\ \order{m^4/Q^2},
\end{align}
where the underbraces collect terms that are $\order{m^2}$, and
the errors induced by approximating $k^+$ are $\order{m^4/Q^2}$.
Therefore, the small components $k^-$ and $\kT$ must be kept
exact to avoid introducing unsuppressed errors.
Implementing this approximation requires another momentum
four-vector $\tilde{k}^\mu$, defined in the Breit frame as
\begin{equation}
\tilde{k} \equiv \parz{\xbj P^+,k^-,\T{k}{} },
\label{e.ktilde}
\end{equation}
so that the replacement $L(k,P) \to L(\tilde{k},P)$ is a good
approximation up to terms suppressed by powers of $\order{m^2/Q^2}$.

Similarly, the internal lines of $J(k')$ are off-shell by
$\order{m^2}$, while the power counting for $k'$ is
\begin{align}
\label{e.kppowercounting}
k' \sim \big( {\order{m^2/Q}},
	      \order{Q},
	      \order{\T{m}{}}
	\big).
\end{align} 
To find a suitable approximation, consider a frame labeled
by ``$*$'', where the outgoing transverse momentum vanishes,
$k'^*_{\rm T} = 0$.
In terms of the Breit frame variables, one has
\begin{align}
k'^* = \left( k^+ + q^+ -\frac{\kT^2}{2(q^-+k^-)}, q^-+k^-, \T{0}{}
       \right),
\end{align}
so that the outgoing parton's virtuality is
\begin{align}
k'^{*\, 2}
& = 2 \left(k^+ + q^+\right) \left(k^- + q^-\right) - \kT^2	\notag\\
& \sim 2 \left(k^+ + q^+\right) q^- - \kT^2
	+ \order{\frac{m^3}{Q}}.
\end{align}
Therefore, the smallest component of $k$, namely $k^-$, can be
neglected in $J(k')$.  To implement this approximation we define the
approximate outgoing momentum four-vector
\begin{equation}
k' \to \tilde{k}' \equiv \left( l^+, q^-, \T{0}{} \right),
\label{e.ktilde'}
\end{equation}
where $l^+ \equiv k^+ - \xbj P^+ + \Tscsq{k}{}/(2q^-)$.
Changing the integration variables from $k^+$ to $l^+$ in
\eref{unappint} gives
\begin{align}
&W^{\mu\nu}(P,q) 
&= \int \frac{\diff{l^+}{} \diff{k^-}{} \diff{^2 \T{k}{}}{} }{(2 \pi)^4}\,
  {\rm Tr} \left[ H^\mu(Q^2) J(l^+) H^{\dagger\nu}(Q^2) L(\tilde{k},P)
	   \right]
+ \order{\frac{m^2}{Q^2}} W^{\mu\nu} \, .
\label{e.unappintt}
\end{align}
The integrations can now be pushed into
separate factors for the target and jet blobs,
\begin{align}
&W^{\mu\nu}(P,q) \no
&=  {\rm Tr} 
\left[ H^\mu(Q^2)
       \parz{\int \frac{\diff{l^+}{}}{2\pi} J(l^+)}
       H^{\nu\dagger}(Q^2)
       \parz{\int \frac{\diff{k^-}{} \diff{^2 \T{k}{}}{}}{(2\pi)^3}
       L(\tilde{k},P)}
\right]
+\ \order{\frac{m^2}{Q^2}} W^{\mu\nu}.
\label{e.unappint2}
\end{align}

To complete the factorization, the jet and target blobs are
decomposed in a basis of Dirac matrices,
\begin{subequations}
\label{e.JLdec}
\begin{align}
J(l^+) & = \gamma_\mu \Delta^\mu(l^+)
            + \Delta_S(l^+)
	    + \gamma_5 \Delta_P(l^+) 
            + \gamma_5 \gamma_\mu \Delta_A^\mu(l^+)
            + \sigma_{\mu \nu} \Delta_T^{\mu\nu}(l^+),
\label{e.Jdec} \\
L(\tilde{k},P) &= \gamma_\mu \Phi^\mu(\tilde{k},P)
                + \Phi_S(\tilde{k},P)
                + \gamma_5 \Phi_P(\tilde{k},P)
                + \gamma_5 \gamma_\mu \Phi_A^\mu(\tilde{k},P)
                + \sigma_{\mu \nu} \Phi_T^{\mu\nu}(\tilde{k},P),
\label{e.Ldec}
\end{align}
\end{subequations}
in terms of vector, scalar, pseudoscalar, axial vector and tensor
functions.  If we focus only on spin- and azimuthally-independent
cross sections, only the first term in \eref{Jdec} and the first
term in \eref{Ldec} need be kept.
To leading power, only the ``$-$'' component of $\Delta^\mu$ and
only the ``$+$'' component of $\Phi^\mu$ contribute, so that the
jet and target operators can be expanded as
\begin{subequations}
\label{e.JLexpn}
\begin{align}
J(l^+) 
  & = \gamma^+ \Delta^-(l^+)
      + \order{\frac{m^2}{Q^2}} J
      + {\rm (spin \; dep.)}			\notag\\
  & = \frac{\slashed{\hat{k}}'}{4 q^-}
	{\rm Tr} \left[ \gamma^- J(l^+) \right]
      + \order{\frac{m^2}{Q^2}} J
      + {\rm (spin \; dep.)}, 			\\
L(\tilde{k},P) 
  & = \gamma^- \Phi^+(\tilde{k},P)
      + \order{\frac{m^2}{Q^2}} L
      + {\rm (spin \; dep.)}			\notag\\
  &= \frac{\slashed{\hat{k}}}{4 \xn P^+}
	{\rm Tr} \left[ \gamma^+ L(\tilde{k},P) \right] 
      + \order{\frac{m^2}{Q^2}} L
      + {\rm (spin \; dep.)},
\end{align}
\end{subequations}
where the spin-dependent terms are not written explicitly.
Using Eqs.~(\ref{e.JLexpn}), the spin-averaged hadronic tensor is then
\begin{eqnarray}
W^{\mu\nu}(P,q)
&=& \frac{ 2 \pi }{ 2 Q^2}\, 
    {\rm Tr} \left[ H^\mu(Q^2) \slashed{\hat{k}}'
	 	    H^{\dagger\nu}(Q^2) \slashed{\hat{k}}
	     \right]
    \parz{ \int  \frac{\diff{l^+}{}}{2 \pi} 
	 {\rm Tr}\left[\frac{\gamma^-}{2} J(l^+) \right]}
\notag\\ 
& & \times
    \parz{ \int \frac{\diff{k^-}{} \diff{^2 \T{k}{}}{}}{(2\pi)^3} 
    {\rm Tr} \left[ \frac{\gamma^+}{2} L(\tilde{k},P) \right] }
    + \order{\frac{m^2}{Q^2}} W^{\mu\nu}.
\label{e.unappint3}
\end{eqnarray}
Finally, the integration contour for $l^+$ is deformed away from
the $k'$ pole until $l^+ q^-$ is $\order{Q^2}$.  To lowest order in
$\lambda^2$, $J$ can then be replaced by the  massless, on-shell
cut diagram, so the hadronic tensor in \eref{unappint3} becomes
\begin{eqnarray}
W^{\mu\nu}(P,q)
&=& \underbrace{\frac{ 2 \pi}{ 2 Q^2} \, {\rm Tr} 
  \left[ H^\mu(Q^2) \slashed{\hat{k}}'
 	 H^{\dagger\nu}(Q^2) \slashed{\hat{k}}
  \right]}_{\mathcal{H}^{\mu\nu}(Q^2)}\;
  \underbrace{\parz{\int\frac{\diff{k^-}{} \diff{^2 \T{k}{}}{}}{{ (2\pi)^4}}
  {\rm Tr} \left[ \frac{\gamma^+}{2} L(\tilde{k},P)
	   \right]}}_{f(\xbj)}
\notag \\
& & +\ \order{\frac{m^2}{Q^2}} W^{\mu\nu}.
\label{e.unappint4}
\end{eqnarray}
This is the standard factorized hadronic tensor.
The hard scattering factor $\mathcal{H}^{\mu\nu}(Q^2)$ contains the
short-distance [$\order{Q^2}$] physics, and the parton distribution
$f(\xbj)$ contains large-distance [$\order{m^2}$] physics associated
with the initial bound state.  The transition from \eref{unappint} to
\eref{unappint4} is represented graphically in \fref{handbag}(a)--(c).

From the hadronic tensor, one recovers the structure functions
in the collinear (parton model) approximation,
\begin{align}
F_i(\xbj,Q^2) = \mathcal{H}_i(Q^2) \; f(\xbj)
    + \order{\frac{m^2}{Q^2}},\ \ \ \ \ i=1,2,
\label{e.unappint5}
\end{align}
where 
\begin{equation}
\mathcal{H}_i(Q^2) 
  \equiv {\rm P}_i^{\mu\nu}\,\frac{ 2 \pi}{ 2 Q^2} {\rm Tr} 
  \left[ H_\mu(Q^2)\, \slashed{\hat{k'}}\,
	 H^\dagger_\nu(Q^2)\, \slashed{\hat{k}}
  \right].
\label{e.H12}
\end{equation}
At leading order, $H^\mu(Q^2) = \gamma^\mu$,
so that the projected hard functions in \eref{H12} become
\begin{subequations}
\label{e.H12final}
\begin{align}
\mathcal{H}_{1}(Q^2)
&= 2 \pi, \\
\mathcal{H}_{2}(Q^2)
&= 4 \pi \frac{Q^2 \xbj \left(Q^2-\mhad^2 \xbj^2\right)}
	 {\left(Q^2 + \mhad^2 \xbj^2\right)^2}	\notag\\
&= 4 \pi \xbj \left( 1 + \order{\frac{M^2 \xbj^2}{Q^2}}
	  \right).
\end{align}
\end{subequations}
The hadronic tensor in \eref{hadtens} is often defined with an overall 
$1/(4 \pi)$. Including this in \eref{unappint5} produces the familiar 
$F_1 = f(\xbj)/2$ and $F_2 = \xbj f(\xbj)$ result of the parton model.

In the limit of large $Q$ and at fixed $\xbj$, the graphs in
Figs.~\ref{f.basicmodel}(B)--(C) are suppressed by powers of $m/Q$,
and the structure function in the factorized approximation comes
entirely from the contribution in \fref{basicmodel}(A).
The graphical topology is a specific instance of the handbag
diagram in \fref{handbag}(c).

The PDF $f(\xbj)$, which describes the lower blob in
\fref{handbag}(a) in the factorized approximation, is 
\begin{eqnarray}
f(\xbj)
&=& \int\frac{\diff{k^-}{} \diff{^2 \T{k}{}}{}}{{ (2\pi)^4}}
    \parz{\frac{1}{\tilde{k}^2 - \mquark^2}}^2\,
    {\rm Tr}
	\left[\frac{\gamma^+}{2}  (\tilde{\slashed{k}} + \mquark)
	      (\slashed{P} + \mhad) (\tilde{\slashed{k}} + \mquark)
	\right]
\notag\\
& & \hspace*{5cm} \times\,
    (2 \pi)\, \delta_+\parz{(P-\tilde{k})^2-\mgluon^2} \, .
\label{e.pm_pdf}
\end{eqnarray}
The on-shell $\delta$-function eliminates the integration over
$k^-$, giving
\begin{align}
k^- &= -\frac{\xbj
	      \left[ \Tscsq{k}{} + \mgluon^2 + (\xbj-1) \mhad^2 \right]}
	     {\sqrt{2} Q (1-\xbj)},
\label{e.kminapp}
\end{align}
and the parton virtuality becomes
\begin{align}
\tilde{k}^2
&= -\frac{\Tscsq{k}{}+\xbj \left[ \mgluon^2 + (\xbj-1)\mhad^2 \right]}
	 {1-\xbj}.
\label{e.ksqd}
\end{align}
Finally, the $\kT$-unintegrated functions ${\cal F}_{1,2}$
defined in Eqs.~(\ref{e.F12kTdef}) are given, in the collinear
factorization approximation, by
\begin{align}
\label{e.calFcoll}
{\cal F}_1(\xbj,Q^2,\kT^2)
&= {\cal F}_2(\xbj,Q^2,\kT^2)\,
 =\,
     \frac{(1-\xbj) \big[\kT^2 + (\mquark +\xbj \mhad )^2\big]}
	  {\big[\kT^2 + \xbj \mgluon^2
		+ (1-\xbj)\, \mquark^2 + \xbj (\xbj-1) \mhad^2
	   \big]^2}.
\end{align}
These structure functions only depend on $\xbj$ and $\kT^2$
and are independent of $Q^2$, as would be anticipated for
the parton model approximation.
The equality ${\cal F}_1 = {\cal F}_2$ is a version of the
Callan-Gross relation \cite{Callan:1969uq}, but for the
unintegrated structure functions.
Note that the parton virtuality $\tilde{k}^2$ in Eq.~\eqref{e.ksqd}
in the PDF is an approximation to the true parton virtuality.

To develop intuition about the approximations just made on the
parton momentum, it is useful to Taylor expand the exact $k^+$,
$k^-$ and $k^2$ from~\erefs{kplusminus}{Delta} through the first
several powers of $m^2/Q^2$,
\begin{align}
\xi ={}& \xbj 
  \left[ 1 + \frac{\Tscsq{k}{} +  \mquark^2-\xbj^2 \mhad^2 }{Q^2} 
  \right.\,  
         \nonumber \\ 
  {}&\qquad\left. 
         - \frac{\xbj^3 \mhad^2 \left(k_T^2+\mquark^2\right)
         +\xbj \left(k_T^2+\mquark^2\right) 
          \left(k_T^2+\mgluon^2-\mhad^2\right)
         -2 \mhad^4 \xbj^4 (\xbj - 1)}{Q^4 \left(\xbj-1\right)} 
  \right] 
  \nonumber \\ 
  {}& \qquad + \order{\frac{m^6}{Q^6}}\, , 
%  {}& \qquad + \order{\left( \frac{m}{Q}\right)^6}\, , 
\label{e.plusexp} \\
& \nonumber \\
k^- ={}& -\frac{\xn}{Q \sqrt{2}} 
  \left[ \frac{\Tscsq{k}{} + \mgluon^2 + (\xn-1) \mhad^2}{1-\xn} 
        -\frac{\xn \left(\Tscsq{k}{}+\mquark^2\right) 
               \left(\Tscsq{k}{}+\mgluon^2\right)}
              {Q^2 (\xn-1)^2} 
   \right]
   \nonumber \\
   {}& \qquad + \order{m \cdot \frac{m^5}{Q^5}}\, , 
%   {}& \qquad + \order{m \left( \frac{m}{Q}\right)^5}\, , 
\label{e.minexp} \\
& \nonumber \\
k^2 ={}& 
  -\frac{\Tscsq{k}{}+\xn \left[\mgluon^2+(\xn-1) \mhad^2\right]}{1-\xn} 
  \nonumber \\
  {}&\qquad
    -\frac{\xn \left(\Tscsq{k}{}+\mquark^2\right) 
      \left( \vphantom{\frac{\frac{M}{M}}{M}} 
             \Tscsq{k}{}
	   + \left[\mgluon+(\xn-1) \mhad\right]
             \left[\mgluon-(\xn-1) \mhad\right]
      \right)}{Q^2 (\xn-1)^2} 
    \nonumber \\
  {}& \qquad + \order{m^2 \cdot \frac{m^4}{Q^4}}\, .
%  {}& \qquad + \order{m^2 \left( \frac{m}{Q}\right)^4}\, .
\label{e.sqexp}
\end{align}
Here we have expressed $\xi$ in terms of $\xbj$ because the leading
power contribution to $\xi$ is conventionally written as $\xbj$.
The lowest non-vanishing powers in \erefs{minexp}{sqexp} match
\erefs{kminapp}{ksqd}, respectively, confirming that the
approximations leading up to Eq.~\eqref{e.calFcoll} are valid for
sufficiently large $Q$.  For $k^-$ and $k^2$, it is more convenient
to maintain expressions in terms of $\xn$.  Of course, $\xn$ may be
replaced everywhere here by $\xbj$ without changing the validity of
the expressions.

The formula for the $\order{\lambda^2}$ PDF in \eref{pm_pdf} could
also have been obtained directly from the operator definition of
the collinear PDF, calculated in the scalar diquark field theory.
The definition of the PDF emerges automatically from the constraints
of factorization. This is an important aspect of the steps above,
and is a key of factorization derivations.

%%%%%%%%%%%%%%%%%%%%%%%%%%%%%%%%%%%%%%%%%%%%%%%%%%%%%%%%%%%%%%%%%%%%%%%%
\section{Exact and factorized structure functions: A comparison}
\label{s.pheno}

In this section we compare DIS structure functions in the exact
calculation of \sref{exact} with the corresponding calculations
in the factorization approximation of \sref{factapp}.
We restrict consideration to unintegrated structure functions,
differential in $\kT$.
This permits a direct examination of the impact of the approximations 
from the previous section point-by-point in transverse momentum.
Exact kinematics involve sensitivity to all components of parton
momentum, including parton virtuality, so the notion of factorization
with a collinear PDF will not apply to the exact case.
However, the terms in a direct $m^2/Q^2$ expansion of the exact
result can hint at ways to correct the collinear picture.

The power counting in Eq.~\eqref{e.powercounting}, with $m^2 \ll Q^2$,
must be reasonably well satisfied for the steps of the previous section
to constitute a good approximation.  Namely, the magnitude of the quark
virtuality $|k^2|$ must be small relative to the hard scale $Q^2$.
While the distribution of $k^2$ in an isolated proton is an intrinsic
property of the bound state, the range of $k^2$ probed in a DIS
collision is sensitive to external kinematical parameters like
$\xbj$ and $\mhad$.  Therefore, the validity of the $|k^2| \ll Q^2$
assumption also depends on external kinematics.

To make this clear, one may directly examine the behavior
of~\erefs{kplusminus}{Delta} in various limiting cases.
For example, consider fixed $Q^2$ and the limit of $\xn \to 1$.
The $\pm$ components of $k$ are then
\begin{subequations}
\begin{align}
& k^+ \to \frac{Q}{\sqrt{2}} 
          \parz{1 + \frac{\mquark^2 - \mgluon^2}{\mhad^2 + Q^2}}
          + \order{|1-\xn|},
\label{e.kpex} \\
& k^- \to -\frac{1}{2 \sqrt{2} Q} \parz{ 
          Q^2 - \mhad^2 
          + \frac{  (\mhad^2 + Q^2)
                    (2 \Tscsq{k}{} + \mgluon^2 + \mquark^2)
                 }{\mgluon^2 - \mquark^2} } 
          + \order{|1-\xn|} 
\label{e.kmex} \, .
\end{align}
\end{subequations}
Next taking the large-$Q^2$ limit, the quark virtuality becomes
\begin{align}
& \lim_{m/Q \to 0}\ \lim_{\xn \to 1} \;\; k^2= -\frac{Q^2}{2}
  \left(
    1 + \frac{2\Tscsq{k}{}+\mquark^2+\mgluon^2}{\mgluon^2-\mquark^2}
  \right).
\label{e.k2ex}
\end{align}
The typical value of $-k^2$ is therefore of order $Q^2$ in the
simultaneous limits of large $\xn$ and large $Q$.
[From Eq.~\eqref{e.sqexp}, this remains true if the order of the
limits is reversed.]
The increasing size of $|k^2|$ with increasing $\xbj$ is a symptom
of parton kinematics becoming non-collinear.  As $\xn$ becomes very
large, it eventually becomes questionable whether an interpretation
in terms of universal collinear parton densities is possible.
We will return to this discussion in \sref{purkin}.

%======================================================================%
\subsection{Values for $\mquark$ and $\mgluon$}
\label{s.paramvals}

To proceed with numerical calculations, we must return to the
discussion in \sref{analogy} regarding choices for $\mquark$
and $\mgluon$.  In QCD, the mass of the target remnant will tend
to grow with energy and $Q^2$, so the choice of $\mgluon$ requires
greater care.  Lower bounds on $\mgluon$ can be obtained from
elementary kinematic considerations.  Since the invariant mass of
the final state system cannot be less than that of the lowest
baryon state, namely the nucleon, then
\begin{equation}
W^2(\xbj,Q) = (\spectator + \jet)^2  > \mhad^2 \, . 
\label{e.Wine}
\end{equation}
Working in the rest frame of the quark--diquark system,
\begin{equation}
\mhad - \mquark < \mgluon \leq W(\xbj,Q) - \mquark \, .
\label{e.massineq}
\end{equation}
This constrains $\mgluon$ to lie in a band whose width depends
on $\xbj$ and $Q$, with the range decreasing as $\xbj \to 1$.

We are interested in the numerical effects of the factorization
approximations for some selected fixed values of $k^2$.
However, $k^2$ is determined by external kinematics
and the field theory parameters $\mquark$ and $\mgluon$. 
Therefore, we will choose $\mgluon$ on a case-by-case basis to
ensure specific values of $k^2$ designed to test power counting
assumptions for reasonable $k^2$.  The relationship between $k^2$
and $\mgluon$ depends on other kinematic parameters, so we will
need to choose a new $\mgluon$ for each kinematical scenario
in order to keep $k^2$ fixed.
To see this, note that for fixed $\xbj$ and large $Q^2$,
the relationship between $\mgluon$ and $k^2$ is
\begin{equation}
\mgluon^2
\approx (1-\xbj) \left( M^2 + \frac{|k^2|}{\xbj} \right)\, .
\label{e.mstrans}
\end{equation}
For different $\xbj$, $\mgluon$ must be modified if $k^2$ is to
remain fixed.  In the next section we will use the exact
relationship between $\mquark$, $\mgluon$, $k^2$ and $\Tsc{k}{}$ to
choose specific values for $\mgluon$ and $\mquark$ so that $|k^2|$
is no greater than several hundred~MeV at small~$\Tsc{k}{}$.

If the actual typical $\Tsc{k}{}$, $k^2$, and $\mquark$ are
clustered around a range of very small values, then collinear
factorization might be satisfied with very high accuracy even
for relatively small $Q$.
However, phenomenological studies of transverse momentum dependence
in semi-inclusive DIS restrict typical $\Tsc{k}{}$-widths to
$\approx500$--800~MeV
\cite{Feynman:1978dt, Anselmino:2013lza, Signori:2013mda},
while model-based estimates suggest
$\langle \Tsc{k}{} \rangle \, \approx \, 300$~MeV
\cite{Georgi:1976ve}. (See also Ref.~\cite{Thomas:2001kw} and references therein.)
Thus, the values  we choose for $\mquark$ and $|k^2|$ (or $\mgluon$)
cannot be simultaneously much less than about 300~MeV without
creating tension with measurements of transverse momentum
dependence in semi-inclusive DIS.
Also, Eq.~\eqref{e.massineq} means that $\mgluon$ cannot be
much less than $\mhad$ if $\mquark$ is small.  
Therefore, we will choose combinations of $\mgluon$ and $\mquark$
such that $|k^2|$ is several hundred~MeV, $\mquark$ is in the
vicinity of $\mquark \approx 300$~MeV, and the peak of the
transverse momentum distribution is not greater than 300~MeV.
[This peak location is somewhat small relative to the above examples
from phenomenology; this will ensure that we underestimate
$\order{\Tscsq{k}{}/Q^2}$ kinematical errors to the collinear
factorization formula.]
The details of the resulting example calculations are discussed
in the following.

%======================================================================%
\subsection{Which power corrections are most important?}
\label{s.tmcs}

In the canonical factorization approximations of \sref{factapp},
there are four independent types of neglected power-suppressed terms:
\begin{subequations}
\begin{align}
\sim\ \frac{\mquark^2}{Q^2}\, ; \, & \qquad {\rm Type-A} \label{e.typea} \\
\sim\ \frac{k^2}{Q^2}\, ; \, & \qquad {\rm Type-B} \label{e.typeb} \\
\sim\ \frac{\Tscsq{k}{}}{Q^2}\, ; \, & \qquad {\rm Type-C} \label{e.typec} \\
\sim\ \frac{M^2}{Q^2}\, . \, & \qquad {\rm Type-D} \label{e.typed} 
\end{align}
\end{subequations}
For the purposes of power counting, we use $k^2$ as the independent
variable for Type--B corrections in place of $\mgluon^2$.
Of course, beyond leading power-law corrections, these suppression
factors come in combinations.  For
example, the $\sim \order{m^6/Q^6}$ power corrections include terms
proportional to
\begin{eqnarray}
\frac{k^2}{Q^2} \times \frac{\Tscsq{k}{}}{Q^2} \times \frac{M^2}{Q^2} \, .
\label{eq:othercorrections}
\end{eqnarray}
Therefore, it is not generally meaningful to address Type--D
suppressed corrections independently of Type--B and Type--C
suppressed corrections.
Effects from $\mhad^2/Q^2$ in higher powers are sensitive
to the range of $k^2$.

Still, it is possible in principle that corrections
suppressed by exactly one type of factor in \erefs{typea}{typed}
alone might be important.  For example, it is reasonable to
speculate that terms with \emph{only} a Type--D suppression
may be large, whereas terms with any of Type--A through Type--C
suppressions are negligible.
Now that the exact and factorized calculations of the structure
functions in the quark--diquark theory are available to us, we can test
the feasibility of such an approximation directly by examining
the relative importance of Type--A through Type--C corrections
as compared with pure Type--D corrections.
When corrections from isolated $\mhad/Q$ terms are useful, the quality
of the approximations from \sref{factapp} should nonetheless be nearly
independent of the exact values of $\kT$, $\mquark$ and $k^2$,
so long as they lie within a reasonable range.
If, however, small variations in $\kT$, $\mquark$ or $k^2$ produce
large changes in the quality of the factorization approximation,
then target mass corrections
from terms like Eq.~\eqref{eq:othercorrections} are too large to
ignore, and it is unlikely that isolated $\mhad/Q$ corrections
alone can improve accuracy.

%......................................................................%
\begin{figure*}
\centering
\includegraphics[width=\textwidth]{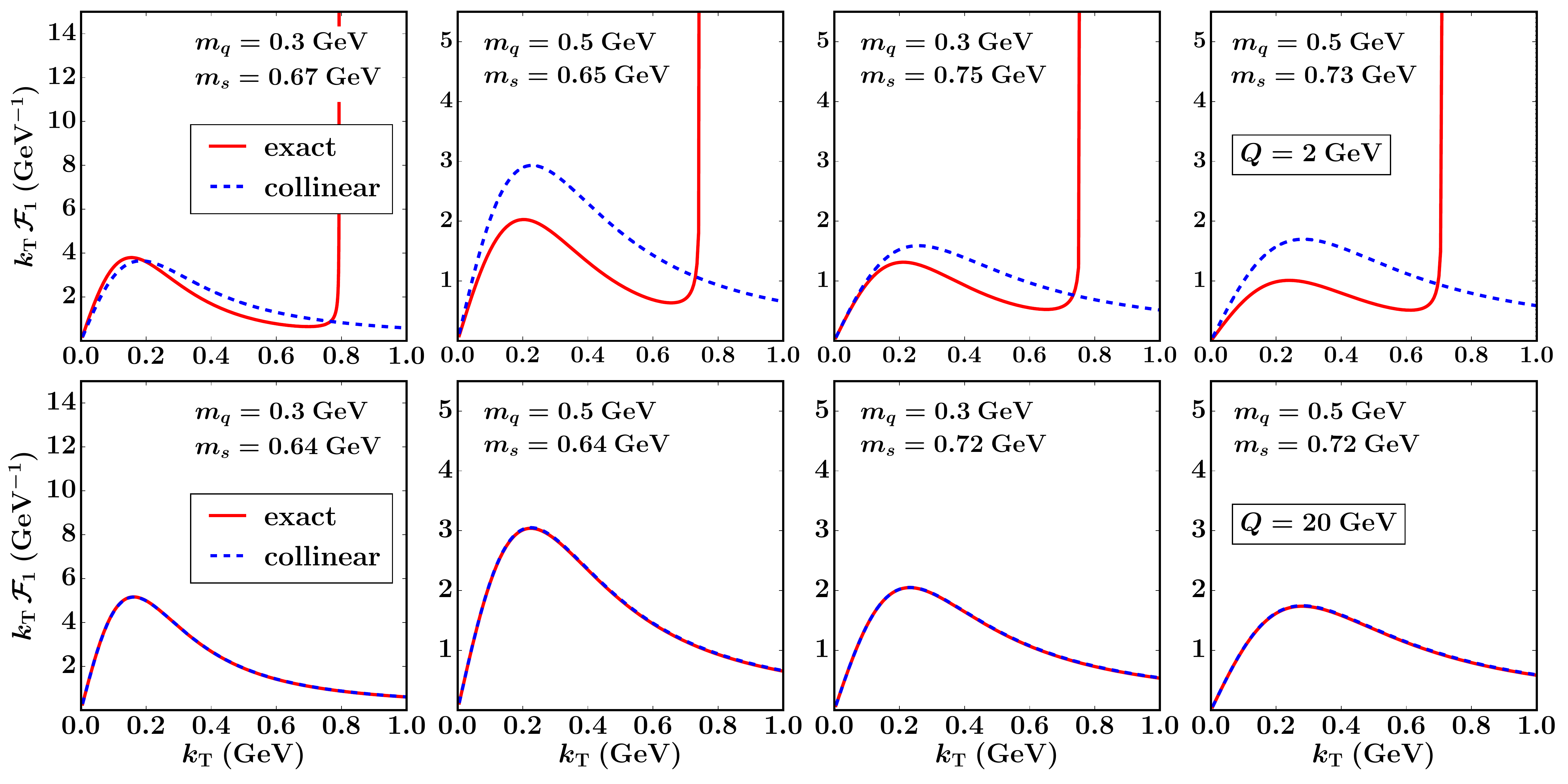}
\caption{
  The unintegrated structure function $\Tsc{k}{} \mathcal{F}_1$ for
  $\xbj = 0.6$ and $Q = 2$~GeV (top row) and $Q = 20$~GeV (bottom row),
  for different values of $\mquark$ and $\mgluon$ calculated using
  both the exact expressions (solid red curves) and the canonical
  collinear factorization approximation (dashed blue curves). The choices of $\mgluon$ are to 
  fix $k^2$ at the values discussed in \sref{paramvals}.
  At the higher $Q$ value the collinear calculation is almost
  indistinguishable from the exact, while at the lower $Q$ value
  the exact calculation diverges as it approaches the kinematical
  upper limit of $\Tsc{k}{}$.
}
\label{f.Q2}
\end{figure*}
%......................................................................%

To illustrate the numerical dependence of the structure functions on
the mass parameters $\mquark$ and $\mgluon$, we show in \fref{Q2}
the unintegrated $\mathcal{F}_1\parz{\xn,Q^2,\kT^2}$ structure
function, weighted by $\Tsc{k}{}$, as a function of $\Tsc{k}{}$.
(The results for the $\mathcal{F}_2$ structure function are
qualitatively similar, and do not alter our conclusions.)
We emphasize that these plots correspond to the $k^-$ solution in \eref{kminus} for which $\left|k^2\right|$ may be small enough to yield parton model kinematics.  The other solution is dealt with in the $\order{\lambda^2}$ hard part. The kinematics are chosen to be representative of typical values
relevant to large-$\xbj$ studies at modern accelerator facilities,
$\xbj = 0.6$ for $Q = 2$~GeV, which corresponds to $W \approx 2$~GeV,
and a higher $Q$ value, $Q = 20$~GeV, characteristic of the deep
scaling region.
For the quark mass we take $\mquark = 0.3$ and 0.5~GeV, while the
values for the diquark mass $\mgluon$ are chosen to ensure that
the quark virtuality
     $v \equiv \sqrt{-k^2} = 300$~MeV or 500~MeV at $\Tsc{k}{}=0$. 
These values are chosen to be consistent with the kinematical
constraints discussed in \sref{paramvals} and, as seen in \fref{Q2},
they produce distributions peaked at $\Tsc{k}{}$ slightly less than
$\approx 300$~MeV.
For the exact calculation, there is an integrable kinematical
square root divergence at $\Tsc{k}{} = \ktmax$ that is an artifact
of our simplification to a $2 \to 2$ process.  All graphs from
\fref{basicmodel} are included now, as required for an
$\order{\lambda^2}$ treatment without kinematical approximations.
Note that with exact kinematics it is now only the sum of the graphs
in \fref{basicmodel} that is gauge invariant.

At the higher $Q$ value in \fref{Q2} (bottom row), the factorized
structure function is almost indistinguishable from the exact result.
This validates that the approximate and exact calculations
match in the large-$Q$ limit, even for $\Tsc{k}{} \gtrsim 1$~GeV.
By contrast, for the lower $Q$ value in \fref{Q2} (top row), the
exact calculation shows a clear deviation from the factorization
approximation, both in size and shape.  It is clear that
if corrections of order $\sim 10\%$ are important, then the roles
of Type-A through Type-C corrections need to be considered on the
same footing with Type-D corrections.  The top row of \fref{Q2} shows
that the quality of the collinear factorization approximations for
$Q \sim$~few~GeV is indeed sensitive to the exact values of $k^2$
and $m_q$, whereas the applicability of the collinear factorization
paradigm assumes independence of these nonperturbative parameters.

Even for the large $Q$ value in \fref{Q2}, the shape of the
$\Tsc{k}{}$ distribution is sensitive to the precise values of
$\mquark$ and $\mgluon$, with the unintegrated structure function
diverging for small values of $\Tsc{k}{}$ as $\mquark$ and
$\mgluon \to 0$.  This is to be expected because the $\Tsc{k}{}$
dependence near $\Tsc{k}{} \approx 0$ is determined by the
nonperturbative physics that regulates the infrared limit in
the hadron wave function.
More relevant is that the approximation errors are vanishingly
small at $\Tsc{k}{} < 1$~GeV and large $Q$, independently of
$\mgluon$ and $\mquark$, as long as they lie within a
reasonable range as discussed in \sref{paramvals}.

Note also that the incoming quark virtuality $k^2$ is forced by
kinematics to decrease to large negative values with increasing
$\Tsc{k}{}$.  This is illustrated in \fref{Q2k2}, which shows the quark
virtuality $v$ as a function of $\Tsc{k}{}$ for fixed $\xbj=0.6$
and $Q=2$ and 20~GeV. The exact and approximate results for $v$
coincide at the high $Q$ value but differ visibly small $\Tsc{k}{}$
and large $\Tsc{k}{}$ for the lower $Q$.
At large $\Tsc{k}{}$, the virtuality becomes linear with $\Tsc{k}{}$,
in accordance with \eref{sqexp} in the $m/Q \to 0$ limit.
Even assuming $v < 1$~GeV for $\Tsc{k}{} < 1$~GeV, the exact value of
$k^2$ (and its dependence on $\Tsc{k}{}$) impacts the shape of the
$\Tsc{k}{}$ distribution and the quality of the usual factorization
approximations.

%......................................................................%
\begin{figure*}
\centering
\includegraphics[width=\textwidth]{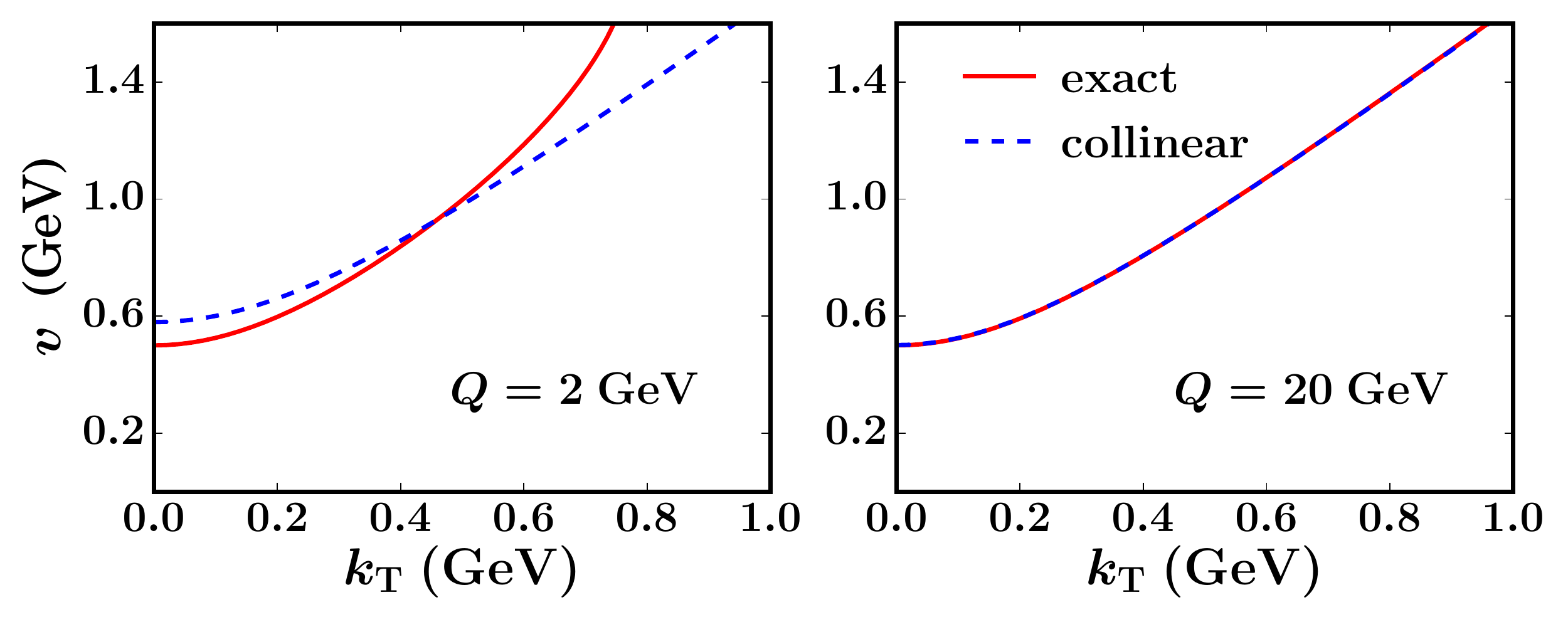}
\caption{
  The dependence of the parton virtuality $v \equiv \sqrt{-k^2}$
  on $\Tsc{k}{}$ evaluated at exact (solid red curves) and
  approximate collinear (dashed blue curves) kinematics,
  for $\xbj=0.6$ at fixed
  $Q=2$~GeV (left panel) and
  $Q=20$~GeV (right panel),
  for quark mass $\mquark=0.3$~GeV and spectator diquark mass
  $\mgluon$ corresponding to $v(\Tsc{k}{}=0)=0.5$~GeV
  (see Table~\ref{t.results}).
}
\label{f.Q2k2}
\end{figure*}
%......................................................................%

%======================================================================%
\subsection{The role of transverse momentum}

The factorization approximations discussed in \sref{factapp}
apply to the limit in which $\kT/Q \sim m/Q \ll 1$.
In QCD, however, there are ultraviolet divergences from the
integrals over transverse momentum in the PDF.
The standard way to deal with this is to renormalize the PDF.

When $Q$ is large, vertex corrections involve $\order{Q^2}$ off-shell
propagators, so the appropriate renormalization scale is $\mu \sim Q$.
By comparison, the kinematics of real gluon emission restrict
``large'' transverse momentum to be
	$\lesssim \order{Q \sqrt{1 - \xbj})}$
[see \eref{ktmax}], so that the corresponding scale is
$\mu \sim Q \sqrt{1 - \xbj}$. 
(In our model calculation, the spectator plays the role
kinematically of a real gluon emission.)
If $\xbj$ is not too large and $Q \gg m$, this mismatch between
real and virtual emissions is not a serious problem because
$\ktmax$ is at least $\order{Q}$ for all graphs.
The collinear parton distribution \eref{pm_pdf} becomes,
schematically,
\begin{equation} 
f(\xbj)\ \propto\ \int_{M_{\rm cut}^2}^{\ktmaxsq \sim Q^2} 
                  \frac{\diff{\Tscsq{k}{}}{}}{\Tscsq{k}{}}\
	 \propto\ \ln \frac{Q^2}{m^2}\, ,
\label{e.ktint}
\end{equation} 
where the lower bound $M_{\rm cut}$ on the integration is to
restrict attention to the large $\Tsc{k}{} \sim Q$ component of
the integration [namely, the contribution to $f(\xbj)$ from the
large-$k_{\rm T}$ region varies logarithmically with $Q^2$].
As long as $\xbj$ is not too large,~\eref{ktint} is consistent
with the corresponding logarithms from virtual loops.
The resulting $\log Q^2$ dependence is the familiar $Q^2$
dependence that arises in the standard DGLAP-type evolution
equations which produce the logarithmic scaling violations of
PDFs~\cite{Gribov:1972ri, Dokshitzer:1977sg, Altarelli:1977zs}.

However, if $\xbj \approx 1 - m^2/Q^2$, then $\ktmax$ is no
greater than $\order{m}$ and the large logarithms of \eref{ktint}
are no longer present.
The ultraviolet divergences from loop integrals still need to be
renormalized at the scale of the virtual photon ($\mu \sim Q$),
so $\ln Q^2$ behavior from loop diagrams remain.  This creates a
mismatch between the renormalization of real and virtual emissions.
In QCD, the mismatch appears in high-order $\alpha_s(Q)$
contributions in the form of uncontrolled large finite parts,
well-known as $\ln(1-\xbj)$ effects that, at a minimum,
need to be resummed to all orders \cite{Sterman:1987aj,
Catani1989, Dokshitzer:2005bf, Manohar:2003vb}.

The small-$\ktmax$ problem is evident in the scalar diquark theory
in \fref{Q2} for the $\xbj = 0.6$ and $Q = 2$~GeV kinematics.
The value of $\Tsc{k}{}$ here approaches its kinematic upper bound at
$\Tsc{k}{} \lesssim 1$~GeV, so the $\Tsc{k}{} \ll Q$ approximation
begins to fail already for $\Tsc{k}{} \sim$~several hundred~MeV.
By contrast, for the higher $Q$ value in \fref{Q2}, the kinematical
upper bound on $\Tsc{k}{}$ lies well above 1~GeV (off the scale of
the graphs).  In QCD, this large $\Tsc{k}{}$ region is generally
describable by perturbative real gluon radiation.

To highlight the trends in $\Tsc{k}{}$ dependence at larger $\xbj$
and moderate $Q$, it is useful to consider the exact $\ktmax$ from
\eref{ktmax} in various limits.  For example, in the limit of small
$m/Q$ with fixed $\xbj$,
\begin{equation}
\ktmax
= \frac{Q}{2} 
  \left[
    \sqrt{\frac{1 - \xbj}{\xbj}}
  - \sqrt{\frac{\xbj}{1-\xbj}} 
    \frac{\left( 2 \mquark^2 + 2 \mgluon^2 - \mhad^2 \right)}{2 Q^2} 
  + \order{\frac{m^4}{Q^4} \parz{\frac{\xbj}{1-\xbj}}^{3/2}}
\right] \, . 
\label{e.ktexp1}
\end{equation} 
This is the fixed-$\xbj$ Bjorken limit applied to $\ktmax$, but a
truncation of the series is liable to be a poor approximation to
$\ktmax$ if $\xbj$ is close to one.  In that limit, it is more
meaningful to Taylor expand first in powers of small $(1 - \xbj)$
with fixed $Q$,
\begin{eqnarray}
\ktmax
&=& \frac{1}{2 M} 
   \sqrt{ (\mquark^2 - \mhad^2)^2 
        + (\mgluon^2 - \mquark^2)^2 
        + (\mgluon^2 - \mhad^2)^2 
        - \mgluon^4 - \mquark^4 - \mhad^4
        }				\notag\\ 
& & +\ \order{(1 - \xbj) \frac{Q^3}{m^2}}\, .
\label{e.ktmaxlargex}
\end{eqnarray}
There is thus a finite and generally nonzero upper bound on $\Tsc{k}{}$
as $\xbj$ becomes large.  Indeed, if the collision is exactly elastic,
$\xbj \to 1$, and \eref{W} requires $\mquark + \mgluon = \mhad$,
which from \eref{ktmaxlargex} gives $\ktmax = 0$.

To quantify errors in the integrations over $\Tsc{k}{}$, we define
the integral over the exact structure function $\mathcal{F}_1$, for
a fixed $\xbj$ and $Q$, between $\Tsc{k}{}=0$ and the kinematic
maximum, $\ktmax$,
\begin{equation}
I(\xbj,Q)
\equiv \int_0^{\ktmax} 
	\diff{\Tsc{k}}{} \; \Tsc{k}{}\,
	\mathcal{F}_1^{\rm exact}(\xbj,Q,\Tsc{k}{})\, .
%	\mathcal{F}_1(\xbj,Q,\Tsc{k}{})\Big|_{\rm exact}\, .
\label{e.Iexact}
\end{equation}
For the analogous calculation in the factorization approximation,
on the other hand, there is no obvious upper bound on the $\Tsc{k}{}$
integration.  In standard treatments, the upper limit, which we denote
by $k_{\rm cut}$, need only be $\order{Q}$, with the exact value
otherwise arbitrary.  Reasonable choices for $k_{\rm cut}$ could be
$\ktmax$ or $Q$, for example.
We define the integral over the structure function in the collinear
approximation as
\begin{equation}
\widehat{I}(\xbj,Q,k_{\rm cut})
\equiv \int_0^{k_{\rm cut}} 
       \diff{\Tsc{k}}{} \; \Tsc{k}{}\,
       \mathcal{F}_1^{\rm approx}(\xbj,Q,\Tsc{k}{})\, .
\label{e.Idef}
\end{equation}
In the limit of large $Q$, as long as
$\order{m} \ll k_{\rm cut} < \order{Q}$, the factorization
approximation should obey
\begin{equation}
\widehat{I}(\xbj,Q,k_{\rm cut}) \approx I(\xbj,Q)\, . 
\label{e.Iratio}
\end{equation}
In QCD, deviations from the equality of $I$ and $\widehat{I}$ are
attributed to higher orders in $\alpha_s(Q)$.  If, however, the ratio
$I/\widehat{I}$ deviates significantly from unity for a range of
reasonable values for $k_{\rm cut}$, the validity of the
collinear factorization approximation begins to become questionable.
Also, $\ktmax$ needs to be $\gtrsim 1$~GeV for gluon radiation effects
to be perturbative.  This is not the case for the $Q=2$~GeV results
in Fig.~\ref{f.Q2}.

%......................................................................%
\begin{table}[t]
\caption{
  Ratio of integrals $I/\widehat{I}$ of exact to collinear
  $\Tsc{k}{}\, \mathcal{F}_1$ structure functions, where
  $I \equiv I(\xbj,Q)$
	[Eq.~(\ref{e.Iexact})] and
  $\widehat{I} \equiv \widehat{I}(\xbj,Q,k_{\rm cut})$
	[Eq.~(\ref{e.Idef})],
  for different values of $\mquark$ and $\mgluon$ as in
  \fref{Q2}, for $\xbj=0.6$ and $Q=2$ and 20~GeV.
  The approximate collinear integral is evaluated for
  $k_{\rm cut} = Q$ and $k_{\rm cut} = \ktmax$.
}
\begin{center}
{\small
\begin{tabular}{c|c|c|c|c|c|c|c|c}	\hline
  & \multicolumn{4}{c|}{$Q=2$~GeV}
  & \multicolumn{4}{c}{$Q=20$~GeV}	\\ \hline
$\mquark$~(GeV)
  & 0.3  & 0.5  & 0.3  & 0.5  
  & 0.3  & 0.5  & 0.3  & 0.5		\\
$\mgluon$~(GeV)
  & ~0.67~  & ~0.65~  & ~0.75~  & ~0.73~ 
  & ~0.64~  & ~0.64~  & ~0.72~  & ~0.72~	\\ \hline
$I/\widehat{I}(\ktmax)$
  & 0.88  & 0.64  & 0.76  & 0.57
  & 1.00  & 1.00  & 1.00  & 1.00	\\
$I/\widehat{I}(Q)$
  & 0.67  & 0.45  & 0.49  & 0.35
  & 0.90  & 0.88  & 0.86  & 0.85	\\ \hline
\end{tabular}
}
\end{center}
\label{t.results}
\end{table}
%......................................................................%

In Table~\ref{t.results} we display the values for $I/\widehat{I}$
using $k_{\rm cut} = \ktmax$ and $k_{\rm cut} = Q$ for the upper
limit on the $\Tsc{k}{}$ integration in $\widehat{I}$, for kinematics
corresponding to \fref{Q2}, namely $\xbj = 0.6$ with $Q = 2$ and 20~GeV.
The values of $\mquark$ and $\mgluon$ are also chosen to be as in
\fref{Q2}, with $\mquark=0.3$ or 0.5~GeV, and $\mgluon$ computed
by fixing the virtuality
$v=0.3$~GeV (smaller $\mgluon$ values, $\sim 0.64$~--~0.67~GeV) or
$v=0.5$~GeV (larger  $\mgluon$ values, $\sim 0.72$~--~0.75~GeV)
at $\Tsc{k}{} = 0$.
For the larger $Q$ value, the results confirm that $I/\widehat{I}$
is approximately unity for $k_{\rm cut}$ between $\ktmax$ and $Q$,
independently of the exact values of $\mquark$ and $\mgluon$,
so long as those values give reasonable $\Tsc{k}{}$ distributions
that peak at $\approx$~few hundred~MeV.
In contrast, for the smaller value of $Q=2$~GeV, the ratio
$I/\widehat{I}$ deviates significantly from unity, and has
stronger dependence on the exact value of $k_{\rm cut}$.
Note that for $Q = 2$~GeV and $\xbj = 0.6$, the maximum transverse
momentum $\ktmax < 1$~GeV, so that the dependence on the
$\Tsc{k}{}$ cutoff likely has its own nonperturbative contributions.

%----------------------------------------------------------------------%
\subsection{Purely kinematic target mass corrections}
\label{s.purkin}

In the context of factorization derivations, the notion of
purely kinematic target mass corrections is unambiguous.
To see this, first return to the factorization approximations
of \sref{factapp}, and assume that for a fixed $\xbj$ and $Q$
the ratio $m^2/Q^2$ is small enough that a power-law expansion
exists and has reasonable convergence.  The first few powers of
the Taylor expansion of momentum components were displayed in
\erefs{plusexp}{sqexp}.
Now assume that, beyond the lowest non-vanishing powers, the only
non-negligible correction terms are those with powers of $M/Q$ alone,
while terms suppressed by higher powers of $\kT/Q$, $\mquark/Q$ or
$\mgluon/Q$ are small.  Upon dropping these, \erefs{plusexp}{sqexp} become
\begin{align}
\xi\ \to\ \xi_{\rm TMC} \equiv{}&
\xbj \left[1 - \frac{\xbj^2 \mhad^2 }{Q^2}
             + \frac{2 \mhad^4 \xbj^4}{Q^4} + \cdots
     \right] = \xn \, , 
\label{e.plusexplim} \\
k^-\ \to\ k^-_{\rm TMC} \equiv{}&
-\frac{\xn \big[ \Tscsq{k}{} + \mgluon^2 + (\xn-1) \mhad^2 \big]}
      {\sqrt{2} Q (1-\xn)} \, ,
\label{e.minexplim} \\
k^2\ \to\ k^2_{\rm TMC} \equiv{}&
-\frac{\Tscsq{k}{}+\xn \big[ \mgluon^2+(\xn-1) \mhad^2 \big]}
      {1-\xn} \, .
\label{e.sqexplim}
\end{align}
Comparing with Eqs.~\eqref{e.kminapp} and~\eqref{e.ksqd} confirms
that using \erefs{plusexplim}{sqexplim} is identical to simply
replacing $\xbj \to \xn$ in the standard collinear parton model
approximation, \eref{calFcoll}.
Indeed, the replacement of $\xn$ by $\xbj$ in \eref{khatall}
was unnecessary for deriving the factorization formula; the steps
leading to the factorized hadronic tensor in \eref{unappint4}
are equally valid if $\xbj$ is replaced everywhere by $\xn$.

There is, therefore, a natural meaning to purely kinematic TMCs:
They are the terms that are kept in the factorization
derivation when all components of external, physical momenta,
such as \erefs{P}{q}, are left unapproximated.
Specifically, purely kinematical TMCs are those that arise from
keeping the minus component of the target momentum $P$, which is
normally approximated to zero, exact in Eq.~\eqref{e.P}.
This automatically results in $\xn$-scaling (often referred to in
the literature as \mbox{``$\xi$-scaling''}, not to be confused
with the $\xi$ variable used for the ``$+$'' component of $k$ here),
as opposed to $\xbj$-scaling.

Power corrections beyond those accounted for in
\erefs{plusexplim}{sqexplim} are associated with $\kT$, $\mquark$
and $k^2$ dependence, and hence are unavoidably coupled to bound
state dynamics that are both nonperturbative and non-collinear
(for $\Tsc{k}{} \sim m$).
For $\xbj > 0.5$, some of the higher power corrections that only
involve $\kT$, $\mquark$ and $\mgluon$ are enhanced by powers of
$\xbj/(1 - \xbj)$ relative to those that only contain $\mhad$
[see \erefs{plusexp}{sqexp} and Eq.~\eqref{e.ktexp1}].
Moreover, the integration over $\kT$ in QCD includes the full range
of nonperturbative transverse momentum between 0 and $\sim 1$~GeV,
and power corrections that depend on $\kT$ can become quite large.
By contrast, purely kinematical TMCs are suppressed at low $\xbj$
by powers of $\xbj^2 \mhad^2/Q^2$.
This suggests that purely kinematical TMCs alone are not likely
to be sufficient in most interesting large-$\xbj$ cases,
except perhaps for unusually heavy hadrons.
In other words, once $Q$ is small enough (or $\xbj$ large enough)
for there to be sensitivity to purely kinematic TMCs, the effects of
other types of power corrections, including non-collinear effects,
already come into play.

To numerically compare purely kinematical TMCs with other power
correction effects, we show the unintegrated structure ${\cal F}_1$
structure function for the exact calculation in \fref{xnxbj},
with $\xbj=0.6$ and $Q=3$~GeV, and with the standard collinear
approximation and with the collinear result corrected for target
mass effects by rescaling $\xbj \to \xn$.
Perhaps surprisingly, in this case the target mass corrected form
deviates further from the exact result than the uncorrected
collinear approximation.
The expectation that purely kinematic TMCs dominate if $\mhad$ is
especially large is borne out in \fref{xnxbj}, where we compare the
various calculations for the case when $\mhad \to 2 \mhad$.
Here, powers of $\mhad/Q$ are large and the expansion in powers of
$\mhad/Q$ certainly fails.  Thus, the $\xbj \to \xn$ replacement
indeed improves the approximation, though there are still significant
errors from the remaining neglected $m/Q$ corrections that are not
particularly small.

%......................................................................%
\begin{figure*}
\centering
\includegraphics[width=\textwidth]{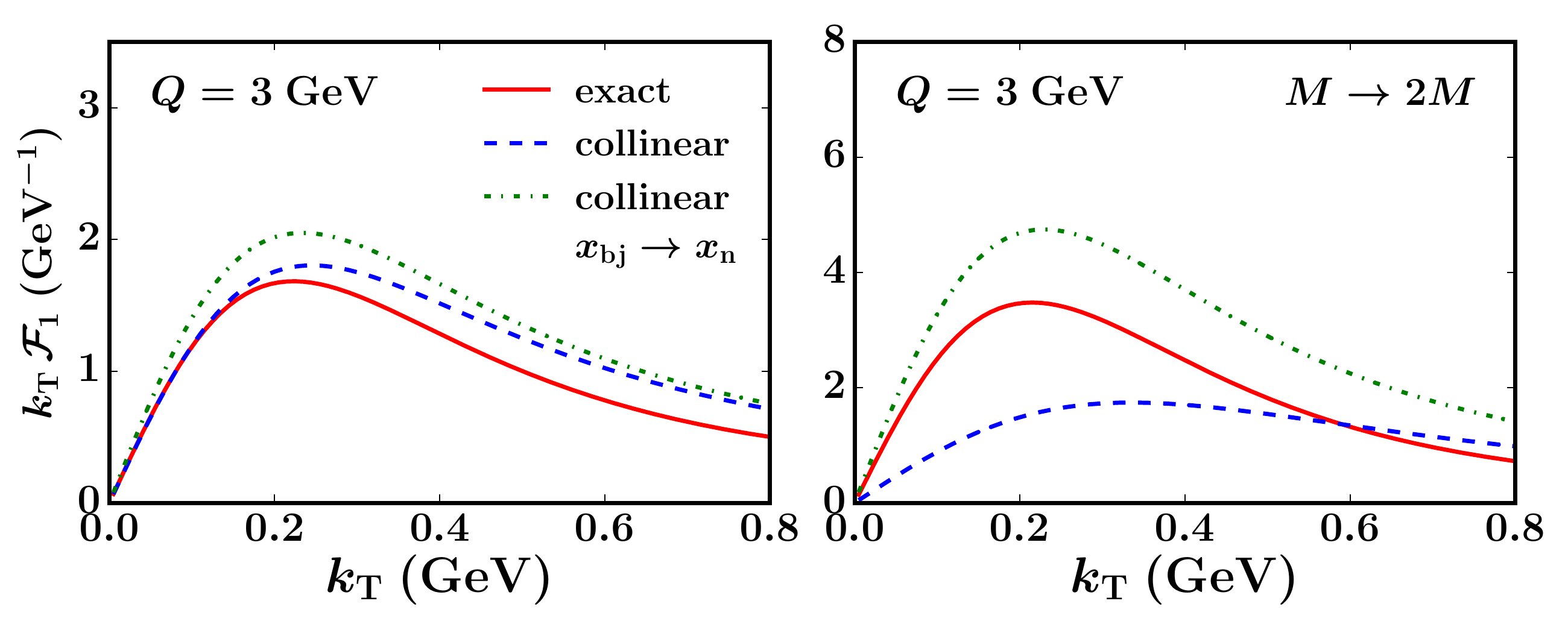}
\caption{
  Unintegrated structure function $\Tsc{k}{} \mathcal{F}_1$ for
  $\xbj = 0.6$ and $Q = 3$~GeV, with quark mass $\mquark=0.3$~GeV and
  virtuality $v=0.5$~GeV for the exact result (solid red curves),
  approximate collinear approximation (dashed blue curves),
  and collinear result with the replacement $\xbj \to \xn$
  (dot-dashed green curves).  The right-hand panel shows the
  results when the nucleon mass increased by a factor of 2.
}
\label{f.xnxbj}
\end{figure*}
%......................................................................%

The phrase ``purely kinematic TMCs" is sometimes used to
characterize the $\order{M^2/Q^2}$ correction terms first derived in
the classic OPE analysis of Georgi and Politzer \cite{Georgi:1976ve}.
The results for the mass corrected structure functions in
Ref.~\cite{Georgi:1976ve} [see Eqs.~(4.19)--(4.22)] differ from those
in \erefs{plusexplim}{sqexplim}, in the form of additional corrections
involving integrals over parton momentum fractions.  These differences
arise because \cite{Georgi:1976ve} imposes the exact constraint
$\tilde{k}^2 = 0$ for quark momentum from the outset.
As explained by Ellis {\it et al.}~\cite{Ellis:1982cd}, the additional
corrections in Ref.~\cite{Georgi:1976ve} originate from the integration
over $\Tsc{k}{}$ when $\tilde{k}^2$ is held fixed at zero.  
In particular, Ref.~\cite{Ellis:1982cd} finds that the unintegrated
structure function must have the functional form [see Eq.~(1.22)]
\begin{equation}
{\cal F}_1\ \sim\ \Phi\parz{\xbj + \frac{\kT^2}{\xbj M^2}}\
		  \theta\left( \xbj(1-\xbj) M^2 - \kT^2 \right).
\label{e.efpform}
\end{equation}
(A similar analysis is given for polarized PDFs in
Ref.~\cite{DAlesio:2009cps}.)
Here, the $\tilde{k}^2 = 0$ condition constrains the behavior of the PDF to all orders in $\xbj \mgluon^2/Q^2$,
$\mquark^2/Q^2$ and $\Tscsq{k}{}/Q^2$.
Furthermore, fixing $\tilde{k}^2 = 0$ removes the ultraviolet
divergences in the integral over $\Tsc{k}{}$ that ultimately gives
rise to the logarithmic behavior characteristic of the DGLAP evolution
equations~\cite{Gribov:1972ri, Dokshitzer:1977sg, Altarelli:1977zs}.
By contrast, factorization derivations impose no constraints on
typical sizes for $\tilde{k}^2$ (recall~\sref{factapp}) inside a PDF, instead
leaving it to be determined by the intrinsic properties of the hadron.

The constraint $\tilde{k}^2 = 0$ in Eq.~\eqref{e.efpform} 
is thus an extra dynamical assumption, and a rather restrictive one.
This is illustrated, for example, by~\fref{Q2k2} and the
discussions in \sref{paramvals}.  In field theory calculations of
a PDF, $k^2$ tends to vary smoothly over a broad range between 0
and $\order{-Q^2}$ (see \fref{Q2k2}), and indeed in an unregulated
integration over $\Tsc{k}{}$, the virtuality $\tilde{k}^2$ diverges.

In practice, the $\tilde{k}^2 = 0$ constraint is rather difficult to
achieve in field theories and realistic models, and it precludes
order-by-order derivations of factorization.  This can be understood
by inspecting \eref{pm_pdf} and noting the distortions to the
$\order{\lambda^2}$ parton distribution that would be necessary
to recover a form like \eref{efpform}.

Figures~\ref{f.Q2}--\ref{f.xnxbj} emphasize that the structure functions
are sensitive to the exact value of $k^2$, including $k^2 \neq 0$.
At a minimum, the higher twist $k^2 \neq 0$ contributions in
Ref.~\cite{Ellis:1982cd} are needed for consistent power counting.
For the above reasons, we will restrict our use of the term
``purely kinematical'' TMCs to what is described in the context
of \erefs{plusexplim}{sqexplim}, namely, only the replacement
$\xbj \to \xn$.

%======================================================================%
\subsection{Help from large $\ln(1 - \xbj)$ resummation}

Beyond leading power in $Q^2$, the integration of the large
transverse momentum in \eref{ktint} actually takes the form
\begin{align}
\int_{M_{\rm cut}^2}^{\ktmax^2}
  \frac{\diff{\Tscsq{k}{}}{}}{\Tscsq{k}{}}{}
&\propto\
  \ln \left[ \frac{Q^2}{M_{\rm cut}^2} \parz{\frac{1 - \xbj}{\xbj} 
          + \frac{\left( \mhad^2 - 2 \mquark^2 - 2 \mgluon^2 \right)}{Q^2}
          + \order{\frac{m^4}{Q^4} \frac{\xbj}{1 - \xbj} } }
        \right] 
\nonumber \\
{}&=\ \ln \frac{Q^2}{M_{\rm cut}^2} + \ln \parz{\frac{1 - \xbj}{\xbj}} 
   + \frac{\xbj \left( \mhad^2 - 2 \mquark^2 - 2 \mgluon^2 \right)}
          {(1 - \xbj) Q^2} \, 
   + \order{\frac{m^4}{Q^4} \frac{\xbj^2}{(1 - \xbj)^2} } \, .  
\label{e.ktint2}
\end{align}
In the region of $\xbj$ where
\begin{equation}
\frac{\xbj m^2}{Q^2}  \ll  1 - \xbj  \ll  1 \, ,
\label{e.largexres}
\end{equation}
the only non-negligible contributions in \eref{ktint2} are the terms
$\ln Q^2$ and $\ln \parz{1 - \xbj}$.  The logarithms of $(1 - \xbj)$
appear at all orders in perturbation theory in collinear factorization,
and much effort has been devoted to methods for resumming them in
collinear perturbative QCD.  It is important to remember, however,
that the usefulness of such methods relies on the condition in
\eref{largexres} being fulfilled.  If hadron mass corrections are
large, for instance when $m^2/Q^2 \sim \alpha_s$, the expansion
\eref{ktint2} may no longer be a useful approximation. In the literal
limit $\xbj \to 1$, it is impossible to fulfill Eq.~\eqref{e.largexres}.

There is of course no obvious sharp boundary between regions where
perturbative \mbox{$\ln(1 - \xbj)$} terms dominate and regions
where $\xbj$ is so large that power corrections dominate or the
power expansion breaks down entirely and \eref{largexres} fails.
In principle, both the logarithmic and power correction effects are
intertwined because they stem from the same underlying physical origin;
the available phase space for final states becomes constricted as
$\xbj \to 1$, and the distinction between logarithmic effects and
subleading power corrections becomes less clear-cut.
For example, it is equally valid to express the large logarithmic
effects in \eref{ktint2} as $\ln(1 - \xbj)$ or $\ln(1 - \xn)$
simply by reorganizing power corrections accordingly.
Thus, incorporating power corrections consistently in perturbative
QCD may entail new techniques in addition to a merging of old ones.

An ideal formalism would smoothly connect a treatment that includes
purely nonperturbative behavior at very large $\xbj$ with resummation
in the limit that the condition in \eref{largexres} holds.
This would be analogous to what occurs with TMD factorization,
where a resummation of $\ln\parz{\Tscsq{q}{}/Q^2}$ holds when
$m \ll \Tsc{q}{} \ll Q$, but nonperturbative intrinsic transverse
momentum dependence contributes when $\Tsc{q}{}$ begins to approach
$m$.  It will be important to explore such effects in future work.

%%%%%%%%%%%%%%%%%%%%%%%%%%%%%%%%%%%%%%%%%%%%%%%%%%%%%%%%%%%%%%%%%%%%%%%%
\section{Summary}
\label{s.conclusions}

Let us conclude by returning to the goals listed at the end of
\sref{intro}.  If it is accepted that the range of values for
$\mquark$ and $\mgluon$ discussed in Secs.~\ref{s.analogy}
and~\ref{s.paramvals} is reasonable, then the results in
\sref{tmcs} indeed imply that all types of power corrections in
\erefs{typea}{typed} are important in the range of $Q \sim 1$~GeV
and $\xbj \gtrsim 0.5$.  For such kinematics, all components of
partonic momentum are potentially non-negligible, and a power
series expansion around the collinear limit may not be sufficient.
Here parton transverse momentum and parton virtuality are as
important as the target mass in determining the size and behavior
of power corrections to collinear factorization.  Moreover, $k^2$
and $\Tsc{k}{}$ are generally not fixed, but rather are correlated
with external kinematic variables such as $\xbj$ and $Q$, and in
principle take a spectrum of values in convolution integrals.

For slightly larger $Q$ and smaller $\xbj$, power corrections will
be smaller but still possibly important.  In all cases, they should
be calculated explicitly in terms of higher twist functions as in
Ref.~\cite{Ellis:1982cd}, or with generalizations of factorization
that take parton kinematics more fully into account.

In the present work, we have placed our analysis of power corrections
in the context of factorization derivations by first reviewing the
canonical collinear factorization approximations for low-order graphs
in \sref{factapp}.  We view this as the appropriate approach to the
treatment of power corrections because collinear factorization is,
fundamentally, the first term in a $1/Q$ expansion, performed
order-by-order in $\alpha_s$ in QCD, or in $\lambda^2$ in the
scalar theory of \eref{lagrangian}.

There are opportunities for extending analyses like the one in
\sref{pheno} and perhaps using them directly for phenomenological
modeling.
In particular, it might be possible to improve constraints on
numerical values for $\mquark$ and $\mgluon$ in a model theory like the scalar Yukawa theory used here by determining if
and how they can be connected to detailed considerations of
nonperturbative physics in QCD.  The values used in this paper were
chosen through a combination of basic kinematical constraints,
extractions of transverse momentum dependent functions,
and mass scales typical of nonperturbative quark models.
In the future, we hope to obtain tighter and more reliable estimates
of the boundary to the factorization collinear regime by appealing
to more sophisticated descriptions of nonperturbative physics.
Including higher-order radiation to model the effects of parton
showering may remove unrealistic features associated with having
a fixed target remnant mass.  Some of these considerations overlap
with the discussions in Ref.~\cite{Boglione:2016bph} of the need
to understand nonperturbative aspects of parton momentum.

We stress that there is in principle a distinction between the
boundary of the collinear kinematics of collinear factorization
and the boundary of the small-$\alpha_s(Q)$ perturbative regime
more broadly.  Thus, an exciting possibility is that there is
a DIS regime at very large $\xbj$ and large $Q$ where collinear
factorization kinematics break down entirely but an alternative
small-$\alpha_s(Q)$ perturbative QCD method applies.
An approach like that of Accardi and Qiu~\cite{Accardi:2008ne},
which takes into account the role of final states in constraining
overall kinematics, is likely needed, but in a form that
incorporates more general noncollinear correlation functions.
Generalizations of PDFs which smoothly map onto the elastic or
exclusive limits may perhaps be appropriate to describe DIS
at very large $\xbj$. 
Models such as the quark--diquark theory used in this work can
provide hints towards more optimal approaches.  The concept of
a virtuality-dependent function, discussed recently by
Radyushkin~\cite{Radyushkin:2016hsy, Radyushkin:2017gjd},
may also play an important role in an improved treatment.
If a particular approximation is valid or useful, it should
be possible to demonstrate the validity of the collinear
approximation in the appropriate limits of \sref{factapp}.
We plan to pursue this in future work.

%%%%%%%%%%%%%%%%%%%%%%%%%%%%%%%%%%%%%%%%%%%%%%%%%%%%%%%%%%%%%%%%%%%%%%%%
\begin{acknowledgments}

We thank A.~Accardi, J.~O.~Gonzalez-Hernandez, S.~Liuti,
A.~Radyushkin, and A.~Rajan for useful discussions. We thank J.~Collins for helpful comments on the text.
This work was supported by the DOE Contract No.~DE-AC05-06OR23177,
under which Jefferson Science Associates, LLC operates Jefferson Lab.
This material is based upon work supported by the U.S. Department
of Energy, Office of Science, Office of Nuclear Physics,
within the framework of the TMD Topical Collaboration.

\end{acknowledgments}

%%%%%%%%%%%%%%%%%%%%%%%%%%%%%%%%%%%%%%%%%%%%%%%%%%%%%%%%%%%%%%%%%%%%%%%%
\appendix

%======================================================================%
\section{Electromagnetic gauge invariance}
\label{a.EMGaugeInv}

In this Appendix we explicitly demonstrate the electromagnetic
gauge invariance of the hadronic tensor $W^{\mu\nu}$ for both the
exact and approximate cases.
Gauge invariance requires $q_\mu W^{\mu\nu}=0$, where $q_\mu$
is the virtual photon momentum.  
In the case of the exact calculation in
\eref{unappintexact}, this means
\begin{equation}
\sum_{j \in \text{Graphs}} q_{\mu}\num_j^{\mu \nu}\, \prop_j\, =\, 0,
\label{e.GaugeInv}
\end{equation}
where $j$ labels the diagrams in \fref{basicmodel}.
To verify \eref{GaugeInv}, we first simplify the contraction for
each diagram individually.

For~\fref{basicmodel}(A),
\begin{subequations}
\begin{align}
&\nonumber q_{\mu}\num_{\rm A}^{\mu\nu}\, \prop_{\rm A}\,
= \frac{{\rm Tr}
   \left[ 
    (\slashed{P} + \mhad)
    (\slashed{k} + \mquark) 
    \slashed{q}
    (\slashed{k} +\slashed{q}+ \mquark) 
    \gamma^\nu
    (\slashed{k} + \mquark) 
   \right]}{(k^2 - \mquark^2)^2}					
\\
&\nonumber =\frac{{\rm Tr}
   \left[ 
    (\slashed{P} + \mhad)
    (\slashed{k} + \mquark) 
    (-(\slashed{k}-\mquark)+\slashed{q}+\slashed{k}-\mquark)
    (\slashed{k} +\slashed{q}+ \mquark) 
    \gamma^\nu
    (\slashed{k} + \mquark) 
   \right]}{(k^2 - \mquark^2)^2}
\\
& =\frac{-{\rm Tr}
   \left[ 
    (\slashed{P} + \mhad)
    (\slashed{k} + \slashed{q} + \mquark) 
    \gamma^\nu
    (\slashed{k} + \mquark) 
   \right]}{(k^2 - \mquark^2)}.
\label{e.GIa}
\end{align}   
For the $1/Q$ suppressed contribution from \fref{basicmodel}(B),
\begin{align}
&\nonumber q_{\mu}\num_{\rm B}^{\mu\nu}\, \prop_{\rm B}\,
= \frac{\, {\rm Tr}
   \left[ 
    (\slashed{P} + \mhad)
    \slashed{q}
    (\slashed{P} +\slashed{q}+ \mhad) 
    (\slashed{k}+\slashed{q} + \mquark) 
    (\slashed{P} +\slashed{q}+ \mhad) 
    \gamma^\nu
   \right]}{\big( (P+q)^2- \mhad^2\big)^2}
\\
&\nonumber =\frac{\, {\rm Tr}
   \left[ 
    (\slashed{P} + \mhad)
    (-(\slashed{P}-\mhad)+\slashed{q}+\slashed{P}-\mhad)
    (\slashed{P} +\slashed{q}+ \mhad) 
    (\slashed{k}+\slashed{q} + \mquark) 
    (\slashed{P} +\slashed{q}+ \mhad) 
    \gamma^\nu
   \right]}{\big( (P+q)^2- \mhad^2\big)^2}
\\
&=\frac{\, {\rm Tr}
   \left[ 
    (\slashed{P} + \mhad)
    (\slashed{k}+\slashed{q} + \mquark) 
    (\slashed{P} +\slashed{q}+ \mhad) 
    \gamma^\nu
   \right]}{\big( (P+q)^2- \mhad^2\big)}.
\label{e.GIb}
\end{align}   
The contribution to \eref{GaugeInv} from the interference
diagram \fref{basicmodel}(C) is 
\begin{align}
&\nonumber q_{\mu}\num_{\rm C}^{\mu\nu}\, \prop_{\rm C}\,
= \frac{{\rm Tr}
   \left[ 
    (\slashed{P} + \mhad)
    (\slashed{k} + \mquark) 
    \slashed{q}
    (\slashed{k} +\slashed{q}+ \mquark) 
    (\slashed{P} +\slashed{q}+ \mhad) 
    \gamma^\nu
   \right]}{(k^2 - \mquark^2)\big( (P+q)^2- \mhad^2\big)}
\\
&\nonumber =\frac{{\rm Tr}
   \left[ 
    (\slashed{P} + \mhad)
    (\slashed{k} + \mquark) 
    (-(\slashed{k}-\mquark)+\slashed{q}+\slashed{k}-\mquark)
    (\slashed{k} +\slashed{q}+ \mquark) 
    (\slashed{P} +\slashed{q}+ \mhad) 
    \gamma^\nu
   \right]}{(k^2 - \mquark^2)\big( (P+q)^2- \mhad^2\big)}
\\
&=\frac{-{\rm Tr}
   \left[ 
    (\slashed{P} + \mhad)
    (\slashed{k}+\slashed{q} + \mquark) 
    (\slashed{P} +\slashed{q}+ \mhad) 
    \gamma^\nu
   \right]}{\big( (P+q)^2- \mhad^2\big)},
\label{e.GIc}
\end{align}
while contribution of the hermitian conjugate of \fref{basicmodel}(C) is
\begin{align}
&\nonumber q_{\mu}\num_{\rm D}^{\mu\nu}\, \prop_{\rm D}\,
= \frac{{\rm Tr}
   \left[\slashed{q} 
    (\slashed{P} +\slashed{q}+ \mhad)
    (\slashed{k} +\slashed{q}+ \mquark) 
     \gamma^\nu
    (\slashed{k} + \mquark) 
    (\slashed{P} + \mhad)  
   \right]}{(k^2 - \mquark^2)\big( (P+q)^2- \mhad^2\big)}
\\
&\nonumber = \frac{{\rm Tr}
   \left[(-(\slashed{P}-\mhad)+\slashed{q}+\slashed{P}-\mhad) 
    (\slashed{P} +\slashed{q}+ \mhad)
    (\slashed{k} +\slashed{q}+ \mquark) 
     \gamma^\nu
    (\slashed{k} + \mquark) 
    (\slashed{P} + \mhad)  
   \right]}{(k^2 - \mquark^2)\big( (P+q)^2- \mhad^2\big)}
\\
&= \frac{{\rm Tr}
   \left[
    (\slashed{k} +\slashed{q}+ \mquark) 
     \gamma^\nu
    (\slashed{k} + \mquark) 
    (\slashed{P} + \mhad)  
   \right]}{(k^2 - \mquark^2)}.
\label{e.GId}
\end{align}
\end{subequations}
Thus,
\begin{equation}
  q_{\mu}\num_{\rm A}^{\mu\nu}\, \prop_{\rm A}
+ q_{\mu}\num_{\rm B}^{\mu\nu}\, \prop_{\rm B}
+ q_{\mu}\num_{\rm C}^{\mu\nu}\, \prop_{\rm C}
+ q_{\mu}\num_{\rm D}^{\mu\nu}\, \prop_{\rm D} = 0 \, .
\end{equation}

In the collinear approximation in \eref{unappint4},
the hadronic tensor is gauge invariant if 
\begin{equation}
q_{\mu}{\rm Tr}\left[ H^\mu(Q^2) \slashed{\hat{k}}'
 	 H^{\dagger\nu}(Q^2) \slashed{\hat{k}}
  \right]=0 \, .
\end{equation}
This is easily verified:
\begin{align}
&\nonumber q_{\mu}{\rm Tr}
  \left[ H^\mu(Q^2) \slashed{\hat{k}}'
         H^{\dagger\nu}(Q^2) \slashed{\hat{k}}
  \right]
= {\rm Tr}\left[\slashed{q}(\slashed{\hat{k}}
                +\slashed{q})\gamma^{\nu}\slashed{\hat{k}}
          \right]
= 4\left( \big(2\hat{k}\cdot{q}-Q^2\big) \hat{k}^{\nu}
         -\hat{k}^2q^{\nu}
   \right)\\
&
= 4 \left( 2\hat{k}^+q^--Q^2 \right) \hat{k}^+
= 4 \left( 2 \frac{Q^2}{\sqrt2} \frac{Q^2}{\sqrt2} - Q^2 \right)
             \frac{Q^2}{\sqrt2}
= 0.
\end{align}
Thus, electromagnetic gauge invariance is validated for both the
exact and approximate, collinear cases.

\newpage
%%%%%%%%%%%%%%%%%%%%%%%%%%%%%%%%%%%%%%%%%%%%%%%%%%%%%%%%%%%%%%%%%%%%%%%%
\bibliography{bibliography}

\providecommand{\noopsort}[1]{}
\begin{thebibliography}{50}
\expandafter\ifx\csname natexlab\endcsname\relax\def\natexlab#1{#1}\fi
\expandafter\ifx\csname bibnamefont\endcsname\relax
  \def\bibnamefont#1{#1}\fi
\expandafter\ifx\csname bibfnamefont\endcsname\relax
  \def\bibfnamefont#1{#1}\fi
\expandafter\ifx\csname citenamefont\endcsname\relax
  \def\citenamefont#1{#1}\fi
\expandafter\ifx\csname url\endcsname\relax
  \def\url#1{\texttt{#1}}\fi
\expandafter\ifx\csname urlprefix\endcsname\relax\def\urlprefix{URL }\fi
\providecommand{\bibinfo}[2]{#2}
\providecommand{\eprint}[2][]{\url{#2}}

\bibitem[{\citenamefont{Collins et~al.}(1988)\citenamefont{Collins, Soper, and
  Sterman}}]{Collins:1988gx}
\bibinfo{author}{\bibfnamefont{J.~C.} \bibnamefont{Collins}},
  \bibinfo{author}{\bibfnamefont{D.~E.} \bibnamefont{Soper}}, \bibnamefont{and}
  \bibinfo{author}{\bibfnamefont{G.}~\bibnamefont{Sterman}},
  \bibinfo{journal}{Adv. Ser. Direct. High Energy Phys.}
  \textbf{\bibinfo{volume}{5}}, \bibinfo{pages}{1} (\bibinfo{year}{1988}),
  \eprint{hep-ph/0409313}.

\bibitem[{\citenamefont{Jimenez-Delgado
  et~al.}(2013)\citenamefont{Jimenez-Delgado, Melnitchouk, and
  Owens}}]{Jimenez-Delgado:2013sma}
\bibinfo{author}{\bibfnamefont{P.}~\bibnamefont{Jimenez-Delgado}},
  \bibinfo{author}{\bibfnamefont{W.}~\bibnamefont{Melnitchouk}},
  \bibnamefont{and} \bibinfo{author}{\bibfnamefont{J.~F.} \bibnamefont{Owens}},
  \bibinfo{journal}{J. Phys.} \textbf{\bibinfo{volume}{G40}},
  \bibinfo{pages}{093102} (\bibinfo{year}{2013}), \eprint{arXiv:1306.6515}.

\bibitem[{\citenamefont{Forte and Watt}(2013)}]{Forte:2013wc}
\bibinfo{author}{\bibfnamefont{S.}~\bibnamefont{Forte}} \bibnamefont{and}
  \bibinfo{author}{\bibfnamefont{G.}~\bibnamefont{Watt}},
  \bibinfo{journal}{Ann. Rev. Nucl. Part. Sci.} \textbf{\bibinfo{volume}{63}},
  \bibinfo{pages}{291} (\bibinfo{year}{2013}), \eprint{arXiv:1301.6754}.

\bibitem[{\citenamefont{Bl{\"u}mlein}(2013)}]{Blumlein:2012bf}
\bibinfo{author}{\bibfnamefont{J.}~\bibnamefont{Bl{\"u}mlein}},
  \bibinfo{journal}{Prog. Part. Nucl. Phys.} \textbf{\bibinfo{volume}{69}},
  \bibinfo{pages}{28} (\bibinfo{year}{2013}), \eprint{arXiv:1208.6087}.

\bibitem[{\citenamefont{Bethke et~al.}(2016)\citenamefont{Bethke, Dissertori,
  and Salam}}]{Bethke:2015etp}
\bibinfo{author}{\bibfnamefont{S.}~\bibnamefont{Bethke}},
  \bibinfo{author}{\bibfnamefont{G.}~\bibnamefont{Dissertori}},
  \bibnamefont{and} \bibinfo{author}{\bibfnamefont{G.~P.} \bibnamefont{Salam}},
  \bibinfo{journal}{EPJ Web Conf.} \textbf{\bibinfo{volume}{120}},
  \bibinfo{pages}{07005} (\bibinfo{year}{2016}).

\bibitem[{\citenamefont{Duke and Roberts}(1980)}]{Duke:1979na}
\bibinfo{author}{\bibfnamefont{D.~W.} \bibnamefont{Duke}} \bibnamefont{and}
  \bibinfo{author}{\bibfnamefont{R.~G.} \bibnamefont{Roberts}},
  \bibinfo{journal}{Nucl. Phys.} \textbf{\bibinfo{volume}{B166}},
  \bibinfo{pages}{243} (\bibinfo{year}{1980}).

\bibitem[{\citenamefont{Devoto et~al.}(1983)\citenamefont{Devoto, Duke, Owens,
  and Roberts}}]{Devoto:1983sh}
\bibinfo{author}{\bibfnamefont{A.}~\bibnamefont{Devoto}},
  \bibinfo{author}{\bibfnamefont{D.~W.} \bibnamefont{Duke}},
  \bibinfo{author}{\bibfnamefont{J.~F.} \bibnamefont{Owens}}, \bibnamefont{and}
  \bibinfo{author}{\bibfnamefont{R.~G.} \bibnamefont{Roberts}},
  \bibinfo{journal}{Phys. Rev.} \textbf{\bibinfo{volume}{D27}},
  \bibinfo{pages}{508} (\bibinfo{year}{1983}).

\bibitem[{\citenamefont{Bloom and Gilman}(1971)}]{Bloom:1971ye}
\bibinfo{author}{\bibfnamefont{E.~D.} \bibnamefont{Bloom}} \bibnamefont{and}
  \bibinfo{author}{\bibfnamefont{F.~J.} \bibnamefont{Gilman}},
  \bibinfo{journal}{Phys. Rev.} \textbf{\bibinfo{volume}{D4}},
  \bibinfo{pages}{2901} (\bibinfo{year}{1971}).

\bibitem[{\citenamefont{De~R{\'u}jula
  et~al.}(1977{\natexlab{a}})\citenamefont{De~R{\'u}jula, Georgi, and
  Politzer}}]{DeRujula:1976baf}
\bibinfo{author}{\bibfnamefont{A.}~\bibnamefont{De~R{\'u}jula}},
  \bibinfo{author}{\bibfnamefont{H.}~\bibnamefont{Georgi}}, \bibnamefont{and}
  \bibinfo{author}{\bibfnamefont{H.~D.} \bibnamefont{Politzer}},
  \bibinfo{journal}{Annals Phys.} \textbf{\bibinfo{volume}{103}},
  \bibinfo{pages}{315} (\bibinfo{year}{1977}{\natexlab{a}}).

\bibitem[{\citenamefont{Poggio et~al.}(1976)\citenamefont{Poggio, Quinn, and
  Weinberg}}]{Poggio:1975af}
\bibinfo{author}{\bibfnamefont{E.~C.} \bibnamefont{Poggio}},
  \bibinfo{author}{\bibfnamefont{H.~R.} \bibnamefont{Quinn}}, \bibnamefont{and}
  \bibinfo{author}{\bibfnamefont{S.}~\bibnamefont{Weinberg}},
  \bibinfo{journal}{Phys. Rev.} \textbf{\bibinfo{volume}{D13}},
  \bibinfo{pages}{1958} (\bibinfo{year}{1976}).

\bibitem[{\citenamefont{Ji and Unrau}(1995)}]{Ji:1994br}
\bibinfo{author}{\bibfnamefont{X.}~\bibnamefont{Ji}} \bibnamefont{and}
  \bibinfo{author}{\bibfnamefont{P.}~\bibnamefont{Unrau}},
  \bibinfo{journal}{Phys. Rev.} \textbf{\bibinfo{volume}{D52}},
  \bibinfo{pages}{72} (\bibinfo{year}{1995}), \eprint{hep-ph/9408317}.

\bibitem[{\citenamefont{Melnitchouk et~al.}(2005)\citenamefont{Melnitchouk,
  Ent, and Keppel}}]{Melnitchouk:2005zr}
\bibinfo{author}{\bibfnamefont{W.}~\bibnamefont{Melnitchouk}},
  \bibinfo{author}{\bibfnamefont{R.}~\bibnamefont{Ent}}, \bibnamefont{and}
  \bibinfo{author}{\bibfnamefont{C.}~\bibnamefont{Keppel}},
  \bibinfo{journal}{Phys. Rept.} \textbf{\bibinfo{volume}{406}},
  \bibinfo{pages}{127} (\bibinfo{year}{2005}), \eprint{hep-ph/0501217}.

\bibitem[{\citenamefont{Courtoy and Liuti}(2013)}]{Courtoy:2013qca}
\bibinfo{author}{\bibfnamefont{A.}~\bibnamefont{Courtoy}} \bibnamefont{and}
  \bibinfo{author}{\bibfnamefont{S.}~\bibnamefont{Liuti}},
  \bibinfo{journal}{Phys. Lett.} \textbf{\bibinfo{volume}{B726}},
  \bibinfo{pages}{320} (\bibinfo{year}{2013}), \eprint{arXiv:1302.4439}.

\bibitem[{\citenamefont{Sterman}(1987)}]{Sterman:1987aj}
\bibinfo{author}{\bibfnamefont{G.}~\bibnamefont{Sterman}},
  \bibinfo{journal}{Nucl. Phys.} \textbf{\bibinfo{volume}{B281}},
  \bibinfo{pages}{310} (\bibinfo{year}{1987}).

\bibitem[{\citenamefont{Catani and Trentadue}(1989)}]{Catani1989}
\bibinfo{author}{\bibfnamefont{S.}~\bibnamefont{Catani}} \bibnamefont{and}
  \bibinfo{author}{\bibfnamefont{L.}~\bibnamefont{Trentadue}},
  \bibinfo{journal}{Nucl. Phys.} \textbf{\bibinfo{volume}{B327}},
  \bibinfo{pages}{323} (\bibinfo{year}{1989}).

\bibitem[{\citenamefont{Dokshitzer et~al.}(2006)\citenamefont{Dokshitzer,
  Marchesini, and Salam}}]{Dokshitzer:2005bf}
\bibinfo{author}{\bibfnamefont{{\relax Yu}.~L.} \bibnamefont{Dokshitzer}},
  \bibinfo{author}{\bibfnamefont{G.}~\bibnamefont{Marchesini}},
  \bibnamefont{and} \bibinfo{author}{\bibfnamefont{G.~P.} \bibnamefont{Salam}},
  \bibinfo{journal}{Phys. Lett.} \textbf{\bibinfo{volume}{B634}},
  \bibinfo{pages}{504} (\bibinfo{year}{2006}), \eprint{hep-ph/0511302}.

\bibitem[{\citenamefont{Manohar}(2003)}]{Manohar:2003vb}
\bibinfo{author}{\bibfnamefont{A.~V.} \bibnamefont{Manohar}},
  \bibinfo{journal}{Phys. Rev.} \textbf{\bibinfo{volume}{D68}},
  \bibinfo{pages}{114019} (\bibinfo{year}{2003}), \eprint{hep-ph/0309176}.

\bibitem[{\citenamefont{Ellis et~al.}(1983)\citenamefont{Ellis, Furmanski, and
  Petronzio}}]{Ellis:1982cd}
\bibinfo{author}{\bibfnamefont{R.~K.} \bibnamefont{Ellis}},
  \bibinfo{author}{\bibfnamefont{W.}~\bibnamefont{Furmanski}},
  \bibnamefont{and}
  \bibinfo{author}{\bibfnamefont{R.}~\bibnamefont{Petronzio}},
  \bibinfo{journal}{Nucl. Phys.} \textbf{\bibinfo{volume}{B212}},
  \bibinfo{pages}{29} (\bibinfo{year}{1983}).

\bibitem[{\citenamefont{Jaffe and Soldate}(1982)}]{Jaffe:1982pm}
\bibinfo{author}{\bibfnamefont{R.~L.} \bibnamefont{Jaffe}} \bibnamefont{and}
  \bibinfo{author}{\bibfnamefont{M.}~\bibnamefont{Soldate}},
  \bibinfo{journal}{Phys. Rev.} \textbf{\bibinfo{volume}{D26}},
  \bibinfo{pages}{49} (\bibinfo{year}{1982}).

\bibitem[{\citenamefont{Schienbein et~al.}(2008)}]{Schienbein:2007gr}
\bibinfo{author}{\bibfnamefont{I.}~\bibnamefont{Schienbein}}
  \bibnamefont{et~al.}, \bibinfo{journal}{J. Phys.}
  \textbf{\bibinfo{volume}{G35}}, \bibinfo{pages}{053101}
  (\bibinfo{year}{2008}), \eprint{arXiv:0709.1775}.

\bibitem[{\citenamefont{Georgi and Politzer}(1976)}]{Georgi:1976ve}
\bibinfo{author}{\bibfnamefont{H.}~\bibnamefont{Georgi}} \bibnamefont{and}
  \bibinfo{author}{\bibfnamefont{H.~D.} \bibnamefont{Politzer}},
  \bibinfo{journal}{Phys. Rev.} \textbf{\bibinfo{volume}{D14}},
  \bibinfo{pages}{1829} (\bibinfo{year}{1976}).

\bibitem[{\citenamefont{Nachtmann}(1973)}]{Nachtmann:1973mr}
\bibinfo{author}{\bibfnamefont{O.}~\bibnamefont{Nachtmann}},
  \bibinfo{journal}{Nucl. Phys.} \textbf{\bibinfo{volume}{B63}},
  \bibinfo{pages}{237} (\bibinfo{year}{1973}).

\bibitem[{\citenamefont{Wilson}(1969)}]{Wilson:1969zs}
\bibinfo{author}{\bibfnamefont{K.~G.} \bibnamefont{Wilson}},
  \bibinfo{journal}{Phys. Rev.} \textbf{\bibinfo{volume}{179}},
  \bibinfo{pages}{1499} (\bibinfo{year}{1969}).

\bibitem[{\citenamefont{Brandt and Preparata}(1971)}]{Brandt:1970kg}
\bibinfo{author}{\bibfnamefont{R.~A.} \bibnamefont{Brandt}} \bibnamefont{and}
  \bibinfo{author}{\bibfnamefont{G.}~\bibnamefont{Preparata}},
  \bibinfo{journal}{Nucl. Phys.} \textbf{\bibinfo{volume}{B27}},
  \bibinfo{pages}{541} (\bibinfo{year}{1971}).

\bibitem[{\citenamefont{Christ et~al.}(1972)\citenamefont{Christ, Hasslacher,
  and Mueller}}]{Christ:1972ms}
\bibinfo{author}{\bibfnamefont{N.~H.} \bibnamefont{Christ}},
  \bibinfo{author}{\bibfnamefont{B.}~\bibnamefont{Hasslacher}},
  \bibnamefont{and} \bibinfo{author}{\bibfnamefont{A.~H.}
  \bibnamefont{Mueller}}, \bibinfo{journal}{Phys. Rev.}
  \textbf{\bibinfo{volume}{D6}}, \bibinfo{pages}{3543} (\bibinfo{year}{1972}).

\bibitem[{\citenamefont{Wandzura and Wilczek}(1977)}]{Wandzura:1977qf}
\bibinfo{author}{\bibfnamefont{S.}~\bibnamefont{Wandzura}} \bibnamefont{and}
  \bibinfo{author}{\bibfnamefont{F.}~\bibnamefont{Wilczek}},
  \bibinfo{journal}{Phys. Lett.} \textbf{\bibinfo{volume}{B72}},
  \bibinfo{pages}{195} (\bibinfo{year}{1977}).

\bibitem[{\citenamefont{Bl{\"u}mlein and Tkabladze}(1999)}]{Blumlein:1998nv}
\bibinfo{author}{\bibfnamefont{J.}~\bibnamefont{Bl{\"u}mlein}}
  \bibnamefont{and}
  \bibinfo{author}{\bibfnamefont{A.}~\bibnamefont{Tkabladze}},
  \bibinfo{journal}{Nucl. Phys.} \textbf{\bibinfo{volume}{B553}},
  \bibinfo{pages}{427} (\bibinfo{year}{1999}), \eprint{hep-ph/9812478}.

\bibitem[{\citenamefont{De~R{\'u}jula
  et~al.}(1977{\natexlab{b}})\citenamefont{De~R{\'u}jula, Georgi, and
  Politzer}}]{DeRujula:1976ih}
\bibinfo{author}{\bibfnamefont{A.}~\bibnamefont{De~R{\'u}jula}},
  \bibinfo{author}{\bibfnamefont{H.}~\bibnamefont{Georgi}}, \bibnamefont{and}
  \bibinfo{author}{\bibfnamefont{H.~D.} \bibnamefont{Politzer}},
  \bibinfo{journal}{Phys. Rev.} \textbf{\bibinfo{volume}{D15}},
  \bibinfo{pages}{2495} (\bibinfo{year}{1977}{\natexlab{b}}).

\bibitem[{\citenamefont{Gross et~al.}(1977)\citenamefont{Gross, Treiman, and
  Wilczek}}]{Gross:1976xt}
\bibinfo{author}{\bibfnamefont{D.~J.} \bibnamefont{Gross}},
  \bibinfo{author}{\bibfnamefont{S.~B.} \bibnamefont{Treiman}},
  \bibnamefont{and} \bibinfo{author}{\bibfnamefont{F.~A.}
  \bibnamefont{Wilczek}}, \bibinfo{journal}{Phys. Rev.}
  \textbf{\bibinfo{volume}{D15}}, \bibinfo{pages}{2486} (\bibinfo{year}{1977}).

\bibitem[{\citenamefont{Bitar et~al.}(1979)\citenamefont{Bitar, Johnson, and
  Tung}}]{Bitar:1978cj}
\bibinfo{author}{\bibfnamefont{K.}~\bibnamefont{Bitar}},
  \bibinfo{author}{\bibfnamefont{P.~W.} \bibnamefont{Johnson}},
  \bibnamefont{and} \bibinfo{author}{\bibfnamefont{W.-K.} \bibnamefont{Tung}},
  \bibinfo{journal}{Phys. Lett.} \textbf{\bibinfo{volume}{B83}},
  \bibinfo{pages}{114} (\bibinfo{year}{1979}).

\bibitem[{\citenamefont{Johnson and Tung}(1979)}]{Johnson:1979ty}
\bibinfo{author}{\bibfnamefont{P.~W.} \bibnamefont{Johnson}} \bibnamefont{and}
  \bibinfo{author}{\bibfnamefont{W.-K.} \bibnamefont{Tung}}, in
  \emph{\bibinfo{booktitle}{{7th International Conference on Neutrinos, Weak
  Interactions and Cosmology -- Neutrino '79, Bergen, Norway, June 18-22,
  1979}}} (\bibinfo{year}{1979}).

\bibitem[{\citenamefont{Steffens et~al.}(2012)\citenamefont{Steffens, Brown,
  Melnitchouk, and Sanches}}]{Steffens:2012jx}
\bibinfo{author}{\bibfnamefont{F.~M.} \bibnamefont{Steffens}},
  \bibinfo{author}{\bibfnamefont{M.~D.} \bibnamefont{Brown}},
  \bibinfo{author}{\bibfnamefont{W.}~\bibnamefont{Melnitchouk}},
  \bibnamefont{and} \bibinfo{author}{\bibfnamefont{S.}~\bibnamefont{Sanches}},
  \bibinfo{journal}{Phys. Rev.} \textbf{\bibinfo{volume}{C86}},
  \bibinfo{pages}{065208} (\bibinfo{year}{2012}), \eprint{arXiv:1210.4398}.

\bibitem[{\citenamefont{Libby and Sterman}(1978)}]{Libby:1978qf}
\bibinfo{author}{\bibfnamefont{S.~B.} \bibnamefont{Libby}} \bibnamefont{and}
  \bibinfo{author}{\bibfnamefont{G.}~\bibnamefont{Sterman}},
  \bibinfo{journal}{Phys. Rev.} \textbf{\bibinfo{volume}{D18}},
  \bibinfo{pages}{3252} (\bibinfo{year}{1978}).

\bibitem[{\citenamefont{Becher et~al.}(2015)\citenamefont{Becher, Broggio, and
  Ferroglia}}]{Becher:2014oda}
\bibinfo{author}{\bibfnamefont{T.}~\bibnamefont{Becher}},
  \bibinfo{author}{\bibfnamefont{A.}~\bibnamefont{Broggio}}, \bibnamefont{and}
  \bibinfo{author}{\bibfnamefont{A.}~\bibnamefont{Ferroglia}},
  \bibinfo{journal}{Lect. Notes Phys.} \textbf{\bibinfo{volume}{896}},
  \bibinfo{pages}{1} (\bibinfo{year}{2015}), \eprint{arXiv:1410.1892}.

\bibitem[{\citenamefont{Collins}(2011)}]{Collins:2011qcdbook}
\bibinfo{author}{\bibfnamefont{J.~C.} \bibnamefont{Collins}},
  \emph{\bibinfo{title}{Foundations of Perturbative QCD}}
  (\bibinfo{publisher}{Cambridge University Press},
  \bibinfo{address}{Cambridge}, \bibinfo{year}{2011}).

\bibitem[{\citenamefont{Feynman et~al.}(1978)\citenamefont{Feynman, Field, and
  Fox}}]{Feynman:1978dt}
\bibinfo{author}{\bibfnamefont{R.~P.} \bibnamefont{Feynman}},
  \bibinfo{author}{\bibfnamefont{R.~D.} \bibnamefont{Field}}, \bibnamefont{and}
  \bibinfo{author}{\bibfnamefont{G.~C.} \bibnamefont{Fox}},
  \bibinfo{journal}{Phys. Rev.} \textbf{\bibinfo{volume}{D18}},
  \bibinfo{pages}{3320} (\bibinfo{year}{1978}).

\bibitem[{\citenamefont{Anselmino et~al.}(2014)\citenamefont{Anselmino,
  Boglione, Gonzalez~H., Melis, and Prokudin}}]{Anselmino:2013lza}
\bibinfo{author}{\bibfnamefont{M.}~\bibnamefont{Anselmino}},
  \bibinfo{author}{\bibfnamefont{M.}~\bibnamefont{Boglione}},
  \bibinfo{author}{\bibfnamefont{J.}~\bibnamefont{Gonzalez~H.}},
  \bibinfo{author}{\bibfnamefont{S.}~\bibnamefont{Melis}}, \bibnamefont{and}
  \bibinfo{author}{\bibfnamefont{A.}~\bibnamefont{Prokudin}},
  \bibinfo{journal}{JHEP} \textbf{\bibinfo{volume}{1404}}, \bibinfo{pages}{005}
  (\bibinfo{year}{2014}), \eprint{arXiv:1312.6261}.

\bibitem[{\citenamefont{Signori et~al.}(2013)\citenamefont{Signori, Bacchetta,
  Radici, and Schnell}}]{Signori:2013mda}
\bibinfo{author}{\bibfnamefont{A.}~\bibnamefont{Signori}},
  \bibinfo{author}{\bibfnamefont{A.}~\bibnamefont{Bacchetta}},
  \bibinfo{author}{\bibfnamefont{M.}~\bibnamefont{Radici}}, \bibnamefont{and}
  \bibinfo{author}{\bibfnamefont{G.}~\bibnamefont{Schnell}},
  \bibinfo{journal}{JHEP} \textbf{\bibinfo{volume}{1311}}, \bibinfo{pages}{194}
  (\bibinfo{year}{2013}), \eprint{arXiv:1309.3507}.

\bibitem[{\citenamefont{Bhaduri}(1988)}]{Bhaduri:1988gc}
\bibinfo{author}{\bibfnamefont{R.~K.} \bibnamefont{Bhaduri}},
  \emph{\bibinfo{title}{{Models of the Nucleon: From Quarks to Soliton}}}
  (\bibinfo{publisher}{Addison-Wesley (1988) (Lecture Notes and Supplements in
  Physics, 22)}, \bibinfo{address}{Redwood City, USA}, \bibinfo{year}{1988}).

\bibitem[{\citenamefont{Bacchetta et~al.}(2007)}]{Bacchetta:2006tn}
\bibinfo{author}{\bibfnamefont{A.}~\bibnamefont{Bacchetta}}
  \bibnamefont{et~al.}, \bibinfo{journal}{JHEP} \textbf{\bibinfo{volume}{02}},
  \bibinfo{pages}{093} (\bibinfo{year}{2007}), \eprint{hep-ph/0611265}.

\bibitem[{\citenamefont{Callan and Gross}(1969)}]{Callan:1969uq}
\bibinfo{author}{\bibfnamefont{C.~G.} \bibnamefont{Callan}} \bibnamefont{and}
  \bibinfo{author}{\bibfnamefont{D.~J.} \bibnamefont{Gross}},
  \bibinfo{journal}{Phys. Rev. Lett.} \textbf{\bibinfo{volume}{22}},
  \bibinfo{pages}{156} (\bibinfo{year}{1969}).

\bibitem[{\citenamefont{Thomas and Weise}(2001)}]{Thomas:2001kw}
\bibinfo{author}{\bibfnamefont{A.~W.} \bibnamefont{Thomas}} \bibnamefont{and}
  \bibinfo{author}{\bibfnamefont{W.}~\bibnamefont{Weise}},
  \emph{\bibinfo{title}{{The Structure of the Nucleon}}}
  (\bibinfo{publisher}{Wiley-VCH}, \bibinfo{address}{Berlin, Germany},
  \bibinfo{year}{2001}).

\bibitem[{\citenamefont{Gribov and Lipatov}(1972)}]{Gribov:1972ri}
\bibinfo{author}{\bibfnamefont{V.~N.} \bibnamefont{Gribov}} \bibnamefont{and}
  \bibinfo{author}{\bibfnamefont{L.~N.} \bibnamefont{Lipatov}},
  \bibinfo{journal}{Sov. J. Nucl. Phys.} \textbf{\bibinfo{volume}{15}},
  \bibinfo{pages}{438} (\bibinfo{year}{1972}).

\bibitem[{\citenamefont{Dokshitzer}(1977)}]{Dokshitzer:1977sg}
\bibinfo{author}{\bibfnamefont{Y.~L.} \bibnamefont{Dokshitzer}},
  \bibinfo{journal}{Sov. Phys. JETP} \textbf{\bibinfo{volume}{46}},
  \bibinfo{pages}{641} (\bibinfo{year}{1977}).

\bibitem[{\citenamefont{Altarelli and Parisi}(1977)}]{Altarelli:1977zs}
\bibinfo{author}{\bibfnamefont{G.}~\bibnamefont{Altarelli}} \bibnamefont{and}
  \bibinfo{author}{\bibfnamefont{G.}~\bibnamefont{Parisi}},
  \bibinfo{journal}{Nucl. Phys.} \textbf{\bibinfo{volume}{B126}},
  \bibinfo{pages}{298} (\bibinfo{year}{1977}).

\bibitem[{\citenamefont{D'Alesio et~al.}(2010)\citenamefont{D'Alesio, Leader,
  and Murgia}}]{DAlesio:2009cps}
\bibinfo{author}{\bibfnamefont{U.}~\bibnamefont{D'Alesio}},
  \bibinfo{author}{\bibfnamefont{E.}~\bibnamefont{Leader}}, \bibnamefont{and}
  \bibinfo{author}{\bibfnamefont{F.}~\bibnamefont{Murgia}},
  \bibinfo{journal}{Phys. Rev.} \textbf{\bibinfo{volume}{D81}},
  \bibinfo{pages}{036010} (\bibinfo{year}{2010}), \eprint{arXiv:0909.5650}.

\bibitem[{\citenamefont{Boglione et~al.}(2016)\citenamefont{Boglione, Collins,
  Gamberg, Gonzalez-Hernandez, Rogers, and Sato}}]{Boglione:2016bph}
\bibinfo{author}{\bibfnamefont{M.}~\bibnamefont{Boglione}},
  \bibinfo{author}{\bibfnamefont{J.}~\bibnamefont{Collins}},
  \bibinfo{author}{\bibfnamefont{L.}~\bibnamefont{Gamberg}},
  \bibinfo{author}{\bibfnamefont{J.~O.} \bibnamefont{Gonzalez-Hernandez}},
  \bibinfo{author}{\bibfnamefont{T.~C.} \bibnamefont{Rogers}},
  \bibnamefont{and} \bibinfo{author}{\bibfnamefont{N.}~\bibnamefont{Sato}}
  (\bibinfo{year}{2016}), \eprint{arXiv:1611.10329}.

\bibitem[{\citenamefont{Accardi and Qiu}(2008)}]{Accardi:2008ne}
\bibinfo{author}{\bibfnamefont{A.}~\bibnamefont{Accardi}} \bibnamefont{and}
  \bibinfo{author}{\bibfnamefont{J.}~\bibnamefont{Qiu}},
  \bibinfo{journal}{JHEP} \textbf{\bibinfo{volume}{07}}, \bibinfo{pages}{090}
  (\bibinfo{year}{2008}), \eprint{arXiv:0805.1496}.

\bibitem[{\citenamefont{Radyushkin}(2016)}]{Radyushkin:2016hsy}
\bibinfo{author}{\bibfnamefont{A.}~\bibnamefont{Radyushkin}}
  (\bibinfo{year}{2016}), \eprint{arXiv:1612.05170}.

\bibitem[{\citenamefont{Radyushkin}(2017)}]{Radyushkin:2017gjd}
\bibinfo{author}{\bibfnamefont{A.}~\bibnamefont{Radyushkin}}
  (\bibinfo{year}{2017}), \eprint{arXiv:1701.02688}.

\end{thebibliography}

\end{document}